\documentclass[prd,amsmath,amssymb,superscriptaddress,onecolumn]{revtex4-2}

\usepackage{bbm}
\usepackage{todonotes}
\usepackage[export]{adjustbox}
\usepackage{graphicx}
\usepackage{epsfig}
\usepackage{braket}

\usepackage{bm}

\usepackage{capt-of}

\usepackage{hyperref}

\numberwithin{equation}{section}

\usepackage[noabbrev]{cleveref} 

\newcommand{\dv}{\text{d}}
  
\usepackage{todonotes}

\begin{document}

\title{\boldmath 
Competition of $\chi_{c}(2P)$ quarkonia and continuum in $e^+ e^- \to e^+ e^- D \bar{D}$}

\author{Izabela Babiarz}
\email{Izabela.Babiarz@ifj.edu.pl}
\affiliation{Institute of Nuclear Physics Polish Academy of Sciences, Radzikowskiego 152, PL-31342 Krak{\'o}w, Poland}

\author{Piotr Lebiedowicz}
\email{Piotr.Lebiedowicz@ifj.edu.pl}
\affiliation{Institute of Nuclear Physics Polish Academy of Sciences, Radzikowskiego 152, PL-31342 Krak{\'o}w, Poland}

\author{Wolfgang Sch\"afer}
\email{Wolfgang.Schafer@ifj.edu.pl}
\affiliation{Institute of Nuclear Physics Polish Academy of Sciences, Radzikowskiego 152, PL-31342 Krak{\'o}w, Poland}

\author{Antoni Szczurek}
\email{Antoni.Szczurek@ifj.edu.pl}
\affiliation{Institute of Nuclear Physics Polish Academy of Sciences, Radzikowskiego 152, PL-31342 Krak{\'o}w, Poland}
\affiliation{Institute of Physics, Faculty of Exact and Technical Sciences, University of Rzesz{\'o}w, 
Pigonia 1, PL-35310 Rzesz{\'o}w, Poland}

\begin{abstract}
We discuss the production of $D \bar{D}$ pairs in $e^+ e^-$ collisions, where $D$ refers to either $D^0$
or $D^+$.
The continuum mechanism with the $t/u$-channel vector-meson $D^*$ exchanges are considered. 
The results of the calculation depend on the parameter of the off-shell form-factor 
for the virtual $D^*$ mesons. 
The $D^* D \gamma$ coupling constants are found from the $D^* \to D \gamma$ decays.
We find relatively large contribution for the $D^0 {\bar D}^0$ channel 
and much smaller contribution in the $D^+ D^-$ channel. 
In the second case we consider also the $D^{\pm}$ exchanges. 
We conclude that the bump at $M_{D^0 {\bar D}^0} = 3.8$~GeV
observed by the Belle and BaBar Collaborations has rather
continuum origin than it corresponds to the broad resonance $\chi_{c0}(3860)$.
We discuss also production of the $\chi_{c2}(3930)$ resonance
which is a candidate for the $\chi_{c2}(2P)$ state.
This state can decay into both $D \bar{D}$ channels, however
the branching fractions are not well known at present.
From a comparison of our model results to the BaBar data we find 
$B(\chi_{c2}(3930) \to D \bar{D}) = 0.58 \pm 0.13$
using the two-photon width $\Gamma_{\gamma \gamma} = 0.544$ keV
(obtained for the Buchm\"uller-Tye potential)
evaluated within the light-front approach (NRQCD limit).
Our finding of $\Gamma_{\gamma \gamma} \times B(\chi_{c2}(3930) \to D \bar{D}) = 0.32 \pm 0.07$ keV
is close to the Belle and BaBar results.
Realistic predictions of the differential distributions in several variables
and integrated cross-sections are given for the Belle II kinematics. 
\end{abstract}



\maketitle

\section{Introduction}
\label{sec:Introdution}

The ground state properties of $P$-wave charmonia 
$\chi_{cJ}$ ($J = 0, 1, 2$) are relatively well known. 
Their decay widths and branching fractions were measured \cite{ParticleDataGroup:2024cfk}.
In contrast, the excited $P$-wave charmonia (their masses, widths,
and branching fractions) are poorly known.
Recently, the production of axial $\chi_{c1}(2P)$ state
in $pp$ \cite{Cisek:2022uqx} 
and $e^+ e^-$ \cite{Babiarz:2023ebe} collisions was discussed.
These studies indicate
that also for the famous $X(3872)$, 
although potentially an exotic state,
the $c \bar{c}$ component can play an important role.
This naturally raises the question of whether similar situation can occur for the $J^P = 0^+$ and $2^+$ states.
A number of observed states are considered as candidates for the first excited $P$-wave charmonia $\chi_{c0,2}(2P)$.
A candidate for the $\chi_{c2}(2P)$ state
($\chi_{c2}(3930)$ in PDG \cite{ParticleDataGroup:2024cfk})
was observed by the Belle Collaboration \cite{Belle:2005rte} 
and later confirmed by the BaBar Collaboration \cite{BaBar:2010jfn} 
in the two-photon fusion process 
$e^+ e^- \to e^+ e^- (\gamma \gamma \to D \bar{D})$
as a resonance 
in the $D^{+} D^{-}$ and $D^0 \bar{D}^0$ invariant mass distributions.
In the $D^0 {\bar D}^0$ channel,
one can also observe a broad enhancement at $M \sim 3.8$~GeV. 
The intriguing problem of searching for $\chi_{c0}(2P)$
was a topic of several works; see \cite{Guo:2012tv,Olsen:2014maa}.
The candidate for the $\chi_{c0}(2P)$ state
is either the $X(3860)$ ($\chi_{c0}(3860)$ in PDG)
or the $X(3915)$ ($\chi_{c0}(3915)$ in PDG).
The $X(3915)$ state is relatively narrow ($\Gamma \simeq 20$~MeV)
and was observed in two-photon fusion in the $\omega J/\psi$ decay mode \cite{BaBar:2012nxg}
as well as in the $B$ meson decays, 
$B \to K J/\psi \omega$ 
(Belle \cite{Belle:2004lle}, BaBar \cite{BaBar:2007vxr}) 
and 
$B^{+} \to D^{+} D^{-} K^{+}$ (LHCb \cite{LHCb:2020pxc}).
The observation of a broad resonance
identified as a new charmonium-like state $X(3860)$
with a preference for $J^{PC} = 0^{++}$ over $2^{++}$
was reported by the Belle Collaboration \cite{Belle:2017egg}  
in the process $e^{+} e^{-} \to J/\psi D \bar{D}$.
In line with the arguments presented there, 
the $X(3860)$ state seems to be a better candidate 
for the $\chi_{c0}(2P)$ charmonium state than the $X(3915)$.
There is still much uncertainty about the above assignments 
for the $\chi_{c0,2}(2P)$ states 
due to limited information on their production 
and decay properties, and therefore further confirmation is required.

The nature of the $\chi_{c0}(3915)$ and $\chi_{c2}(3930)$ charmonium-like states 
is also not well understood.
In Ref.~\cite{Zhou:2015uva} the authors
suggest that the two resonances $X(3915)$ and $X(3930)$
can be regarded as the same $J^{PC} = 2^{++}$ state.
It was shown in \cite{Ortega:2017qmg},
using the framework of a constituent quark model,
that this may be related to the molecular structure for these two states.
Further considerations on the $X(3915)/X(3930)$ puzzle 
can be found in \cite{Baru:2017fgv} 
and in the references therein.
Prompt hadroproduction of the $\chi_{c2}(3930)$ resonance
was observed by the LHCb Collaboration
in the decay modes $\chi_{c2}(3930) \to D^{+} D^{-}$ and $D^0 \bar{D}^0$ 
in proton-proton collisions \cite{LHCb:2019lnr}.
A more recent LHCb amplitude analysis of the process
$B^{+} \to D^{+} D^{-} K^{+}$ finds distinct
$J^{PC} = 0^{++}$ and $2^{++}$ states decaying to $D^{+}D^{-}$ \cite{LHCb:2020pxc}, 
which can be identified as 
the $\chi_{c0}(3915)$ and $\chi_{c2}(3930)$
charmonium states, respectively.

Regarding the reactions $e^+ e^- \to e^+ e^- D \bar{D}$,
where $D = D^{0}$, $D^{+}$,
we are not convinced of the existence of a broad resonance 
($\Gamma \sim 200$~MeV)
at $M \sim 3.85$~GeV
reported by the Belle and BaBar Collaborations. 
The reason being that while an enhancement is observed in the $D^0 {\bar D}^0$ invariant mass distribution,
the effect is not clearly seen in the $D^+ D^-$ channel.
In order to increase the statistics
several groups combined the $D^0 {\bar D}^0$ and $D^+ D^-$ final states and studied the combined data. 
For example,
some of the first descriptions of the combined Belle and BaBar data 
on $\gamma \gamma \to D \bar{D}$ reactions were reported in \cite{Guo:2012tv,Chen:2012wy,Zhou:2015uva}.
More recently, the data-driven analysis of $\gamma \gamma \to D \bar{D}$
performed in a partial-wave dispersive formalism \cite{Deineka:2021aeu},
indicates no evidence for the broad resonance $X^{*}(3860)$, but rather reports a bound state below the $D \bar{D}$ threshold. 
In \cite{Wang:2020elp} a possible bound state of $D \bar{D}$ 
generated by the meson-meson interaction is discussed.
In Ref.~\cite{Ji:2022vdj} the available data in the $D \bar{D}$ and $D_{s}^{+} D_{s}^{-}$ channels 
from both $\gamma \gamma$-fusion reaction and 
$B^{+} \to D^{+}D^{-} K^{+}$, $B^{+} \to D_{s}^{+} D_{s}^{-} K^{+}$ decays
were analyzed considering the $D \bar{D}$-$D_{s} \bar{D}_{s}$-$D^{*} \bar{D}^{*}$-$D_{s}^{*} \bar{D}_{s}^{*}$ coupled channels with scalar and tensor states.

In this paper, we consider two different mechanisms
contributing to the $e^{+} e^{-} \to e^{+} e^{-} D \bar{D}$ reactions. 
The first mechanism is the so-called continuum contribution described by $t$- and $u$-channel meson exchanges and contact terms.
For the $D^0 {\bar D}^0$ final state
we consider the vector-meson $D^{*}(2007)^{0}$ $t/u$-channel exchanges
as the underlying continuum process
while in the $D^+ D^-$ case
we consider the pseudoscalar $D^{\pm}$ $t/u$-channel exchanges plus a contact term.
The second production mechanism proceeds through an intermediate resonance state.
Here, we shall examine only the $\chi_{c0}(3860)$ and $\chi_{c2}(3930)$ resonances, 
assuming that they correspond to $\chi_{c0}(2P)$ and $\chi_{c2}(2P)$, respectively.
We stress that we aim to describe the Belle and BaBar data simultaneously, for both the neutral and charged channels separately.

In this paper we discuss the production of $D \bar{D}$ meson pairs in $e^{+} e^{-}$ collisions, 
concentrating on low $\gamma \gamma$ energies ($W_{\gamma \gamma} < 4.3$~GeV).
In \cite{Luszczak:2011js}, the interested reader may find predictions 
for the exclusive $D \bar{D}$ production 
in ultraperipheral ultrarelativistic heavy-ion collisions (UPCs). 
There, the $\gamma \gamma \to D \bar{D}$ subprocesses 
were estimated for $W_{\gamma \gamma} > 4$~GeV 
in the heavy-quark approximation 
and in the Brodsky-Lepage formalism.
It is also worth noting that the production of open-charm meson pairs in UPCs
is one of the potential opportunities to search for the lightest possible exotic charmonium state; 
see \cite{Sobrinho:2024tre}.

\section{Production of $D \bar{D}$ pairs in $e^+ e^-$ collisions via $\gamma \gamma$ fusion}
\label{sec:Formalism}

\begin{figure}
\includegraphics[width=0.3\textwidth]{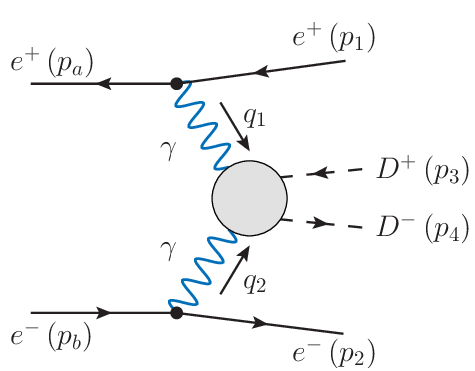}
\quad
\includegraphics[width=0.64\textwidth]{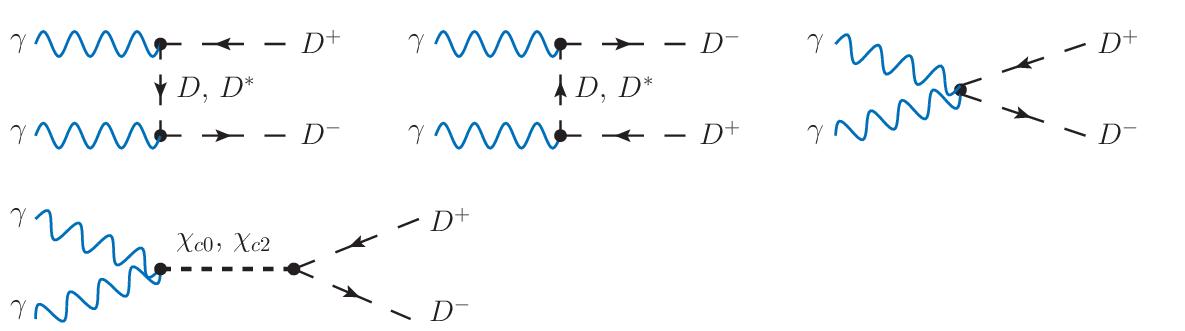}
\caption{Diagrams representing $D^{+}D^{-}$ production in $e^{+}e^{-}$ collisions.
For the process $\gamma \gamma \to D^{+} D^{-}$ we consider 
the $t$- and $u$-channel $D^{\pm}$ and $D^{*}(2010)^{\pm}$ exchanges,
the contact term,
and the exchange of $\chi_{c0,2}$ mesons in the $s$ channel.
The diagrams for the $D^0 {\bar D}^0$ production are similar, 
but in the case of non-resonant production
we consider only the $D^{*}(2007)^{0}$ $t/u$-channel exchanges.}
\label{fig:diagram1}
\end{figure}
In the present studies, we perform  calculations for the considered exclusive
$2 \to 4$ reaction;
see the diagrams shown in Fig.~\ref{fig:diagram1}.
The fully differential cross section reads
\begin{eqnarray}
d \sigma = \frac{1}{2 \sqrt{s (s - 4m_{e}^{2})}} 
\overline{|{\cal M}_{2 \to 4}|^{2}} (2 \pi)^{4}
\delta^{(4)}(p_{1} + p_{2} + p_{3} + p_{4} - p_{a} - p_{b}) 
\prod_{i=1}^{4}
\frac{\dv^{3}p_{i}}{(2\pi)^{3}2p_{i}^{0}}\,.
\label{xs_2to4}
\end{eqnarray}
Here four-momenta of incoming and outgoing particles are denoted by 
$p_{a}, p_{b}, p_{1}, \ldots, p_{4}$,
the energy variable is $s = (p_{a} + p_{b})^{2}$,
$m_{e}$ is the electron mass, and
$\overline{|{\cal M}_{2 \to 4}|^{2}}$ is the amplitude squared averaged over initial
and summed over final lepton polarization states.
To calculate the total cross section one has to perform an eight dimensional phase space integration as is done
for instance for the $pp \to pp \pi^{+} \pi^{-}$ reaction
\cite{Lebiedowicz:2009pj}
and for the $pp \to pp \mu^{+} \mu^{-}$ reaction \cite{Lebiedowicz:2018muq}.
In the present paper, we use the following integration variables:
the transverse momenta of outgoing leptons ($p_{t1}$, $p_{t2}$),
the azimuthal angles of outgoing leptons ($\phi_{1}$, $\phi_{2}$),
the rapidities of the mesons ($y_{3}$, $y_{4}$),
the length and the azimuthal angle of the difference 
of the transverse momenta of the mesons ($p_{m t}$, $\phi_{m}$).

We are interested in the photon-photon fusion production mechanism relevant for small--angle scattering of leptons. The generic $2 \to 4$ amplitude has the form
\begin{eqnarray}
{\cal M}_{\lambda_a \lambda_b \to \lambda_1 \lambda_2 D \bar D} = 
i j^{(1)}_{\alpha}(\lambda_1, \lambda_a) \, 
\frac{- i g^{\alpha \mu}}{q_1^2} \, 
{\cal M}_{\mu \nu}(\gamma^* \gamma^* \to D \bar D) \, 
\frac{-i g^{\nu \beta}}{q_2^2} \, 
i j^{(2)}_\beta(\lambda_2,\lambda_b) \,,
\end{eqnarray}
where $\lambda_{a}, \lambda_{b}, \lambda_{1}, \lambda_{2} \in \{1/2, -1/2 \}$
are the lepton polarizations, and
$q_{1} = p_{a} - p_{1}$, 
$q_{2} = p_{b} - p_{2}$.
Below, we also use the Mandelstam variables
$t_{1} = q_{1}^{2}$, $t_{2} = q_{2}^{2}$, and
$p_{34} = q_{1} + q_{2} = p_{3} + p_{4}$. Furthermore,
\begin{eqnarray}
j^{(1)}_{\alpha}(\lambda_1, \lambda_a)  &=& \bar{v}(p_{1}, \lambda_{1})
\gamma_{\alpha}
v(p_{a}, \lambda_{a}) 
\approx (p_{1} + p_{a})_{\alpha} \,\delta_{\lambda_1 \lambda_a}
\,, \nonumber \\
j^{(2)}_{\beta}(\lambda_2, \lambda_b) &=& 
\bar{u}(p_{2}, \lambda_{2}) 
\gamma_\beta 
u(p_{b}, \lambda_{b})
\approx (p_{2} + p_{b})_{\beta} \,\delta_{\lambda_2 \lambda_b}
\,,
\end{eqnarray}
are the lepton currents.
Let us now collect the amplitudes for the various 
$\gamma^* \gamma^* \to D \bar D$ subprocesses considered by us.

\subsection{Resonant production: $\gamma^* \gamma^* \to \chi_{c2} \to D \bar{D}$}
\label{sec:Formalism_A}
The amplitude for the $\gamma^*(q_1) \gamma^*(q_2) \to \chi_{c2} \to D(p_3) \bar{D}(p_4)$ subprocess is given by
\begin{eqnarray}
{\cal M}_{\mu \nu}(\gamma^{*} \gamma^{*} \to \chi_{c2} \to D \bar{D}) 
&=& 
i\Gamma^{(\gamma^* \gamma^* \to \chi_{c2})}_{\mu \nu \rho \sigma}(q_{1},q_{2}) \;
i\Delta^{(\chi_{c2})\,\rho \sigma, \alpha \beta}(p_{34})\;
i\Gamma^{(\chi_{c2} \to D \bar{D})}_{\alpha \beta}(p_{3},p_{4})\,.
\label{3.1}
\end{eqnarray}

Here, we have introduced the
propagator for a narrow tensor-meson resonance in
the relativistic Breit-Wigner form
\begin{eqnarray}
i\Delta_{\mu \nu, \kappa \lambda}^{(\chi_{c2})}(p_{34})&=&
\frac{i}{p_{34}^{2}-M_{\chi_{c2}}^2+i M_{\chi_{c2}} \Gamma_{\chi_{c2}}}
\left[ 
\frac{1}{2} 
( \hat{g}_{\mu \kappa} \hat{g}_{\nu \lambda}  + \hat{g}_{\mu \lambda} \hat{g}_{\nu \kappa} )
-\frac{1}{3} 
\hat{g}_{\mu \nu} \hat{g}_{\kappa \lambda}
\right] \,, 
\label{3.2}
\end{eqnarray}
where $\hat{g}_{\mu \nu} = -g_{\mu \nu} + p_{34 \mu} p_{34 \nu} / p_{34}^2$.
The total decay width of the $\chi_{c2}(3930)$ resonance and its mass is taken from the PDG~\cite{ParticleDataGroup:2024cfk} as:
\begin{eqnarray}
M_{\chi_{c2}} &=& 3922.5 \pm 1.0~{\rm MeV}\,, \nonumber \\
\Gamma_{\chi_{c2}} &=& 35.2 \pm 2.2~{\rm MeV}\,.
\label{3.3}
\end{eqnarray}

\subsubsection{Two photons vertex: $\gamma^* \gamma^* \to \chi_{c2}$ }

The production of the $\chi_{c2}$ resonance is encoded in the $\gamma^* \gamma^* \to \chi_{c2}$ vertex, which in general depends on five invariant form factors that are functions of the photon virtualities $q_1^2,q_2^2$.
For the corresponding Lorentz structures, see e.g. \cite{Poppe:1986dq,Pascalutsa:2012pr,Babiarz:2020jkh}. 
Here we concentrate on the resonance 
$\chi_{c2}(3930)$ which can be identified 
as the $\chi_{c2}(2P)$ charmonium $c \bar{c}$ state.
We anticipate that the main contribution to the cross section will come from quasireal transversally polarized photons.
In addition, we expect that the helicity-2 component will dominate over the helicity-0 one.
For example in the nonrelativistic quark model, the contribution of the $J_z=0$ polarization of the heavy tensor meson vanishes, see e.g. \cite{Schuler:1997yw}.
We therefore limit ourselves only to helicity-2 component, so that the vertex
can be written as
\begin{equation}
\Gamma_{\mu \nu \kappa \lambda}^{(\gamma^* \gamma^* \to \chi_{c2})}(q_{1},q_{2})=
e^2 
\frac{1}{2} \Big( R_{\mu \kappa}  R_{\nu \lambda}  
                + R_{\nu \kappa}  R_{\mu \lambda}
                - R_{\mu \nu}  R_{\kappa \lambda}  \Big) F_{TT,2}(Q_1^2,Q_2^2)
 \,,
\label{2.1}
\end{equation}
where $e^{2} = 4 \pi \alpha_{\rm em}$,
the photon virtualities
$Q_{1,2}^{2} \equiv -q_{1,2}^{2} \geqslant 0$,
and 
%
\begin{equation}
R_{\mu \nu} \equiv R_{\mu \nu}(q_1,q_2)
= -g_{\mu \nu} + \frac{1}{X}
\Big( (q_{1} \cdot q_{2}) (q_{1 \mu} q_{2 \nu} + q_{1 \nu} q_{2 \mu})
- q_{1}^{2} q_{2 \mu} q_{2 \nu} - q_{2}^{2} q_{1 \mu} q_{1 \nu} \Big)\,,
\label{2.2}
\end{equation}
$X = (q_{1} \cdot q_{2})^{2} - q_{1}^{2} q_{2}^{2}$. 
This vertex function satisfies the relations
\begin{eqnarray}
&&\Gamma_{\mu \nu \kappa \lambda}(q_{1},q_{2})=
\Gamma_{\nu \mu \kappa \lambda}(q_{2},q_{1})\,, \\
&&q_{1}^{\mu}\Gamma_{\mu \nu \kappa \lambda}(q_{1},q_{2})=0\,, \\
&&q_{2}^{\nu}\Gamma_{\mu \nu \kappa \lambda}(q_{1},q_{2})=0\,, \\
&&\Gamma_{\mu \nu \kappa \lambda}(q_{1},q_{2}) g^{\kappa \lambda}=0\,,\\
&&\Gamma_{\mu \nu \kappa \lambda}(q_{1},q_{2}) 
(q_{1} + q_{2})^{\kappa} (q_{1} + q_{2})^{\lambda} = 0\,,
\label{2.3}
\end{eqnarray}
and is also symmetric when exchanging $\kappa \leftrightarrow \lambda$.

The transition form factor will be parametrized as
\begin{eqnarray}
F_{TT,2}(Q_1^2,Q_2^2) = F_{TT,2}(0,0) 
\frac{\Lambda^2}{Q_1^2 + Q_2^2 + \Lambda^2} \,.
\label{2.7}
\end{eqnarray}
In the calculations 
we take $\Lambda = M_{\chi_{c2}(3930)} = 3.9225$~GeV 
which, in principle, is a free parameter.
We would like to mention that $F_{TT,2}(0,0)$,
the form factor at the on-shell point,
is not dimensionless.

Assuming that the tensor $\chi_{c2}(2P)$ state is quarkonium-like 
(pure $c \bar{c}$ state)
one can calculate $F_{TT,2}(0,0)$ in the NRQCD limit 
[see (4.2) of \cite{Babiarz:2024sqw}]
\begin{equation}
F_{TT,2}(0,0) = 8 e_f^2 \sqrt{\frac{3 N_c}{\pi M_{\chi_{c2}}^{3}}} |R_{2P}'(0)|
\label{2.12}
\end{equation}
with $e_f = 2/3$ and $N_{c} = 3$.
The first derivative of the radial wave function at the origin is estimated for several $c \bar{c}$ interaction potential models from the literature, listed in Table~\ref{tab:R0_LFWF} below.
Following the approach in \cite{Babiarz:2020jkh}, we solve Schr\"odinger equation, but for $2P$ state, and get $|R_{2P}'(0)|$ values.
As shown in Table~\ref{tab:R0_LFWF}, there is a considerable spread of results for the different potential models.
\begin{table}[h!]
    \caption{Results of the derivative of the wave function 
    at the origin $|R_{2P}'(0)|$ for $\chi_{c2}(2P)$ state
    obtained from the light front quark model for five different potential model,
    the corresponding values of transition form factor (\ref{2.12}),
    and of two-photon width for the helicity-2 contribution in NRQCD limit.}
    \centering
    \begin{tabular}{l |c |c |c }
    \hline
    \hline
     Potential type &  
     $|R'_{2P}(0)|\, (\rm GeV^{5/2})$ & 
     $|F_{TT,2}(0,0)| \, (\rm GeV)$ &
     $\Gamma_{\gamma \gamma}(\lambda = \pm 2) \,(\rm keV)$ \\
     \hline
    Cornell             & 0.405 & 0.314 & 0.840 \\
    Buchm\"uller-Tye    & 0.326 & 0.253 & 0.544 \\
    harmonic oscillator & 0.275 & 0.213 & 0.387 \\
    logarithmic         & 0.263 & 0.204 & 0.354 \\
    power like          & 0.228 & 0.177 & 0.266 \\
         \hline
         \hline
    \end{tabular}
    \label{tab:R0_LFWF}
\end{table}

In the NRQCD limit, the helicity-0 term vanishes
and the helicity-2 is the dominant one.
Then, $F_{TT,2}(0,0)$ is related to the two-photon decay 
width of the $\chi_{c2}$ state
\begin{eqnarray}
\Gamma(\chi_{c2}(2P) \to \gamma \gamma) =
(4 \pi \alpha_{\rm em})^2 \frac{|F_{TT,2}(0,0)|^2}{80 \pi M_{\chi_{c2}}}\,;
\end{eqnarray}
see the discussion in Sec.~4 of \cite{Babiarz:2024sqw}.
Using $|R_{2P}'(0)| = (0.228 - 0.405)$~GeV$^{5/2}$
we find $\Gamma_{\gamma \gamma}(\lambda = \pm 2) = (0.27 - 0.84)$~keV
assuming the dominance of the helicity-2 term.
Here, the Cornell potential gives the largest value of the two-photon width.
It is worth noting that the BLFQ prediction \cite{Li:2021ejv}
for the two-photon width of $\chi_{c2}(3930)$, 
$\Gamma_{\gamma \gamma} = 0.58(25)$~keV (see Table~I of \cite{Li:2021ejv}),
is most compatible with our result obtained for the Buchm\"uller-Tye potential.

For comparison, a value
$|R_{2P}'(0)|^2 = 0.1767$~GeV$^5$ was found in \cite{Eichten:2019hbb}
computed for the mass $M = 3.9554$~GeV
with $m_{c} = 1.84$~GeV
within the Cornell 
potential.
This value, $|R_{2P}'(0)| \simeq 0.420$~GeV$^{5/2}$,
appears to be larger than our findings.
Nevertheless, it was used 
for the inclusive $\chi_{c1}(3872)$ production \cite{Cisek:2022uqx}
leading to a good description of the proton-proton data
within the uncertainty of the model (different gluon uPDFs).

Alternatively, for the $\gamma^* \gamma^* \to \chi_{c2}$ vertex
one can use the form given in (C25) of \cite{Pascalutsa:2012pr}.
In this case, the helicity-2 component reads
\begin{equation}
\Gamma_{\mu \nu \kappa \lambda}^{\rm PPV}(q_{1},q_{2})=
e^2 
\Big(
R_{\mu \kappa}(q_1,q_2) 
R_{\nu \lambda}(q_1,q_2) 
+
\frac{\hat{s}}{8X}
R_{\mu \nu}(q_1,q_2) 
(q_{1} - q_{2})_{\kappa} (q_{1} - q_{2})_{\lambda} \Big)
\frac{q_{1} \cdot q_{2}}{M_{\chi_{c2}}}
\tilde{F}_{TT,2}(Q_1^2,Q_2^2)
 \,,
\label{2.9}
\end{equation}
where
$\hat{s} = (q_{1} + q_{2})^{2} = 2 (q_{1} \cdot q_{2})
+ q_{1}^{2} + q_{2}^{2}$.
We have $q_{1} \cdot q_{2} = (M_{\chi_{c2}}^{2} + Q_{1}^{2} + Q_{2}^{2})/2$.
Introducing an extra factor $(q_{1} \cdot q_{2})/M_{\chi_{c2}}$ 
the transition form factor $\tilde{F}_{TT,2}(Q_1^2,Q_2^2)$ remains dimensionless.
For the vertex function
(\ref{2.9}) we find the relations
\begin{eqnarray}
&&\Gamma_{\mu \nu \kappa \lambda}^{\rm PPV}(q_{1},q_{2}) g^{\kappa \lambda} \neq 0\,,\\
&&\Gamma_{\mu \nu \kappa \lambda}^{\rm PPV}(q_{1},q_{2}) 
(q_{1} + q_{2})^{\kappa} (q_{1} + q_{2})^{\lambda} \neq 0\,.
\label{2.11}
\end{eqnarray}
In order to get compatibility with the formula given by 
Eq.~(\ref{2.1}), we write
\begin{equation}
\tilde{F}_{TT,2}(Q_1^2,Q_2^2) = 
\frac{M_{\chi_{c2}}}{q_{1} \cdot q_{2}} F_{TT,2}(Q_1^2,Q_2^2)
\,.
\label{2.10}
\end{equation}

The vertices (\ref{2.1}) and (\ref{2.9})
are written for `on-shell' $\chi_{c2}$ meson,
that is for $p_{34}^{2} = M_{\chi_{c2}}^{2}$.
For $p_{34}^{2} \neq M_{\chi_{c2}}^{2}$ 
the form factor $F^{(\chi_{c2})}(p_{34}^{2})$
can be inserted in addition.
%
%
The form factor should be normalized to 
$F^{(\chi_{c2})}(M_{\chi_{c2}}^{2}) = 1$.
In our calculation we put $F^{(\chi_{c2})}(p_{34}^{2}) = 1$.

\subsubsection{$\chi_{c2} \to D \bar{D}$ vertex}

The $\chi_{c2} D \bar{D}$ vertex 
(see Eq.~(3.37) of \cite{Ewerz:2013kda}
for the analogous $f_{2} \pi \pi$ vertex)
can be written as
\begin{eqnarray}
i\Gamma_{\mu \nu}^{(\chi_{c2} \to D \bar{D})}(p_{3},p_{4})=
-i \,\frac{g_{\chi_{c2} D\bar{D}}}{2 M_{0}} 
\left[ (p_{3}-p_{4})_{\mu} (p_{3}-p_{4})_{\nu}
- \frac{1}{4} g_{\mu \nu} (p_{3}-p_{4})^{2} \right] F^{(\chi_{c2})}(p_{34}^{2})
\label{3.4}
\end{eqnarray}
with $M_{0} \equiv 1$~GeV. 
We have 
$\Gamma_{\mu \nu}^{(\chi_{c2} \to D \bar{D})}(p_{3},p_{4}) g^{\mu \nu} = 0$.
Here we can assume the form factor for the off-shell $\chi_{c2}$ meson. 
In our calculation we put $F^{(\chi_{c2})}(p_{34}^{2}) = 1$.

The decay width of $\chi_{c2} \to D^{+} D^{-}$ is given by
\begin{eqnarray}
\Gamma(\chi_{c2} \to D^{+} D^{-}) = 
\frac{M_{\chi_{c2}}}{480 \pi}\,
|g_{\chi_{c2} D\bar{D}}|^{2} 
\left( \frac{M_{\chi_{c2}}}{M_{0}} \right)^{2}
\left( 1-\frac{4 m_{D}^{2}}{M_{\chi_{c2}}^{2}} \right)^{5/2}\,.
\label{3.6}
\end{eqnarray}
We take $m_{D} = m_{D^{+}}$.
The $\chi_{c2}(3930)$ has isospin $I = 0$. 
Assuming isospin invariance in the decay, we get
\begin{equation}
\Gamma(\chi_{c2} \to D^{+} D^{-}) = 
\Gamma(\chi_{c2} \to D^{0} \bar{D}^{0}) = 
\frac{1}{2}\Gamma(\chi_{c2} \to D \bar{D}) \,.
\label{3.7}
\end{equation}

We noted that the most significant partial widths for 
$\chi_{c2}(2P)$ are $\Gamma(\chi_{c2} \to D \bar{D})$ 
and $\Gamma(\chi_{c2} \to D \bar{D}^{*})$.
Different calculations in the literature 
support the fact that
the total width of $\chi_{c2}(2P)$ 
is largely saturated by the $D \bar{D}$ mode.
For instance, 
in Eq.~(24) and Fig.~24 of \cite{Wilson:2023anv} one can find
$\Gamma(\chi_{c2} \to D \bar{D}) \approx 26 (12)$~MeV and
$\Gamma(\chi_{c2} \to D \bar{D}^{*}) \approx 22 (14)$~MeV.
In Table~I of \cite{Gui:2018rvv}
one can find
$\Gamma(\chi_{c2} \to D \bar{D}) = 24$~MeV and
$\Gamma(\chi_{c2} \to D \bar{D}^{*}) = 13.6-16.6$~MeV (depending on model).
In Table~IV of \cite{Wang:2022dfd}, 
the following values corresponding to $M_{\chi_{c2}(2P)} = 3930$~MeV are stated
$\Gamma(\chi_{c2} \to D \bar{D}) = 20.4$~MeV and
$\Gamma(\chi_{c2} \to D \bar{D}^{*}) = 6.9$~MeV.
On the other hand, the authors of \cite{Ortega:2017qmg,Man:2024mvl} 
found a different decomposition of partial widths.

Thus, for the estimate, we take 
$\Gamma(\chi_{c2} \to D \bar{D}) = 14 - 26$~MeV,
and using 
(\ref{3.3}),
(\ref{3.6}), and (\ref{3.7})
we obtain
$|g_{\chi_{c2} D\bar{D}}| = 8.34 - 11.37$.
With $\Gamma_{\chi_{c2}} = 35.2$~MeV (\ref{Gamma_gamgam})
we get the branching fraction in the range
$B(\chi_{c2} \to D \bar{D}) = 0.40 - 0.74$.

\subsection{Resonant production: $\gamma^* \gamma^* \to \chi_{c0} \to D \bar D$}\label{sec:Formalism_chic0}

Here we write formulas for the production of the scalar
$\chi_{c0}(2P)$ state assumed by us to have the mass of $\chi_{c0}(3860)$.
The total decay width of the $\chi_{c0}(3860)$ resonance 
and its mass given in the PDG~\cite{ParticleDataGroup:2024cfk} is:
\begin{eqnarray}
M_{\chi_{c0}} &=& 3862^{+50}_{-35}~{\rm MeV}\,, \nonumber \\
\Gamma_{\chi_{c0}} &=& 200^{+180}_{-110}~{\rm MeV}\,.
\label{4.0_pdg}
\end{eqnarray}
In the calculations, we also consider another scenario:
\begin{eqnarray}
M_{\chi_{c0}} &=& 3862~{\rm MeV}\,, \nonumber \\
\Gamma_{\chi_{c0}} &=& 50~{\rm MeV}\,.
\label{4.0_fit}
\end{eqnarray}
%
The narrower total width of $\chi_{c0}(3860)$ seems to be consistent 
with BaBar data for $D^+D^-$ channel;
see Fig.~\ref{fig:M34_BaBar}(b) below.
Similar value is given in Table~VI of \cite{Eichten:2004uh} 
(see $2^{3}P_{0}$ state).

The amplitude for the $\gamma^*(q_1) \gamma^*(q_2) \to \chi_{c0} \to D(p_3) \bar{D}(p_4)$ 
subprocess is given by
\begin{eqnarray}
{\cal M}_{\mu \nu}(\gamma^{*} \gamma^{*} \to \chi_{c0} \to D \bar{D}) 
=
i\Gamma^{(\gamma^* \gamma^* \to \chi_{c0})}_{\mu \nu}(q_{1},q_{2}) \,
i\Delta^{(\chi_{c0})}(p_{34})\,
i\Gamma^{(\chi_{c0} \to D \bar{D})}(p_{3},p_{4})\,.
\label{4.1}
\end{eqnarray}
The scalar-meson propagator is parametrized as
\begin{eqnarray}
i\Delta^{(\chi_{c0})}(p_{34})=
\frac{i}{p_{34}^{2}-M_{\chi_{c0}}^2+i M_{\chi_{c0}} \Gamma_{\chi_{c0}}}
\label{4.1a}
\end{eqnarray}
with a constant decay width.

The $\gamma^* \gamma^* \to \chi_{c0}$ vertex,
limiting only to the transverse component,
can be written as
\begin{equation}
\Gamma_{\mu \nu}^{(\gamma^* \gamma^* \to \chi_{c0})}(q_{1},q_{2}) =
-e^2 
R_{\mu \nu}
F_{TT}(Q_1^2,Q_2^2)\,.
\label{4.2}
\end{equation}
Here, $R_{\mu \nu}$ is defined by (\ref{2.2}) and the form factor $F_{TT}$ is parametrized as
\begin{eqnarray}
F_{TT}(Q_1^2,Q_2^2) = F_{TT}(0,0) 
\frac{\Lambda^2}{Q_1^2 + Q_2^2 + \Lambda^2} \,, \quad \Lambda = M_{\chi_{c0}}\,.
\label{2.3a}
\end{eqnarray}
Assuming that the state is quarkonium-like one can calculate
$F_{TT}(0,0)$ in the NRQCD limit as 
[see (4.11) of \cite{Babiarz:2020jkh}]
\begin{equation}
F_{TT}(0,0) = e_f^2 \sqrt{N_c} \frac{12}{\sqrt{\pi}}
\frac{R_{2P}'(0)}{M_{\chi_{c0}}^{3/2}}\,.
\label{4.4}
\end{equation}
The two-photon decay width for $\chi_{c0}(2P) \to \gamma \gamma$
is given by
\begin{equation}
\Gamma(\chi_{c0}(2P) \to \gamma \gamma) =
\frac{\pi \alpha_{\rm em}^2}{M_{\chi_{c0}}} |F_{TT}(0,0)|^2 \,.
\label{4.5} 
\end{equation}
In Table~\ref{tab:chic0_2P} we listed our estimates 
for $|F_{TT}(0,0)|$ and $\Gamma_{\gamma \gamma}$ 
for the $\chi_{c0}(2P)$ state.
These values are larger than for $\chi_{c2}(2P)$ given in 
Table~\ref{tab:R0_LFWF}.
It is worth noting that the BLFQ approach \cite{Li:2021ejv}
for the two-photon width of $\chi_{c0}(2P)$
predicts
$\Gamma_{\gamma \gamma} = 0.68(22)$~keV.
\begin{table}[h!]
    \caption{Results for $\chi_{c0}(2P)$ state with $M_{\chi_{c0}} = 3862$~MeV
    obtained from the light front quark model for different potential models in NRQCD limit.}
    \centering
    \begin{tabular}{l |c |c |c }
    \hline
    \hline
     Potential type &  
     $|R'_{2P}(0)|\, (\rm GeV^{5/2})$ & 
     $|F_{TT}(0,0)| \, (\rm GeV)$ &
     $\Gamma_{\gamma \gamma} \,(\rm keV)$ \\
     \hline
    Cornell             & 0.405 & 0.278 & 3.351 \\
    Buchm\"uller-Tye    & 0.326 & 0.224 & 2.171 \\
    harmonic oscillator & 0.275 & 0.189 & 1.545 \\
    logarithmic         & 0.263 & 0.181 & 1.413 \\
    power like          & 0.228 & 0.157 & 1.062 \\
          \hline
         \hline
    \end{tabular}
    \label{tab:chic0_2P}
\end{table}

For the $\chi_{c0} D \bar{D}$ vertex we have ($M_{0} \equiv 1$~GeV)
\begin{eqnarray}
i\Gamma^{(\chi_{c0} \to D \bar{D})}(p_{3},p_{4})=
i g_{\chi_{c0} D\bar{D}} M_{0} \,
F^{(\chi_{c0})}(p_{34}^{2})\,,
\label{4.6}
\end{eqnarray}
where the coupling constant $g_{\chi_{c0} D\bar{D}}$ is related to the partial
decay width of the $\chi_{c0}$ meson
\begin{eqnarray}
\Gamma{(\chi_{c0} \to D \bar{D})}=
\frac{M_{0}^{2}}{16 \pi M_{\chi_{c0}}} |g_{\chi_{c0} D\bar{D}}|^{2} 
\left( 1-\frac{4 m_{D}^{2}}{M_{\chi_{c0}}^{2}} \right)^{1/2}\,.
\label{4.7}
\end{eqnarray}
For the isoscalar $\chi_{c0}$ we have
\begin{equation}
\Gamma(\chi_{c0} \to D^{+} D^{-}) = 
\Gamma(\chi_{c0} \to D^{0} \bar{D}^{0}) = 
\frac{1}{2}\Gamma(\chi_{c0} \to D \bar{D}) \,.
\label{4.8}
\end{equation}
In (\ref{4.2}) and (\ref{4.6}) we can assume the form factor for off-shell $\chi_{c0}$ meson.
In our calculations we put $F^{(\chi_{c0})}(p_{34}^{2}) = 1$.

There is considerable uncertainty in the literature regarding 
the estimation of open-charm strong decay width.
For instance, in Sec.~III~A and Table~II of \cite{Wang:2022dfd}
one can find $\Gamma(\chi_{c0} \to D \bar{D}) = 21.0(16.4)$~MeV 
for $M_{\chi_{c0}} = 3836.8(3862.0)$~MeV,
which is rather narrow.
Similar values for the open-charm strong decay width are predicted 
in \cite{Gui:2018rvv} within
the linear (screened) potential model, and the results are:
$\Gamma(\chi_{c0} \to D \bar{D}) = 22 (28)$~MeV 
for $M_{\chi_{c0}} = 3869(3848)$~MeV.
In the analysis of \cite{Man:2024mvl} 
the width is predicted to be 16.6~MeV,
which is in agreement with the results of \cite{Gui:2018rvv,Wang:2022dfd}
where the adopted decay model is also
the $^{3}P_{0}$ pair creation model.
In \cite{Zhou:2017dwj}, a smaller width about 11~MeV was obtained,
while the analysis of \cite{Yu:2017bsj} in a $^{3}P_{0}$ model
obtained a large width of 110--180~MeV
[which is consistent with (\ref{4.0_pdg})].
The recent analysis of \cite{Gao:2025tob} 
predicts the strong decay width of 73~MeV for $\chi_{c0}(2P)$ state.
This suggest that further precise measurement of the $\chi_{c0}(3860)$ state parameters
and more theoretical studies of the $\chi_{c0}(2P)$ state are still needed.

Taking $M_{\chi_{c0}} = 3862$~MeV,
$\Gamma(\chi_{c0} \to D \bar{D}) = 16.4$~MeV,
and using Eqs.~(\ref{4.7}), (\ref{4.8}) we obtain
$|g_{\chi_{c0} D\bar{D}}| = 2.52$.
This value for $g_{\chi_{c0} D\bar{D}}$ is used in our calculations presented in Sec.~\ref{sec:Results}.
Considering the Buchm\"uller-Tye potential from Table~\ref{tab:chic0_2P}
and assuming 
the branching fraction
$B(\chi_{c0} \to D \bar{D}) = 
\Gamma(\chi_{c0} \to D \bar{D})/\Gamma_{\chi_{c0}}
=
16.4~{\rm MeV}/ 50.0~{\rm MeV} = 0.328$
we get
\begin{equation}
\Gamma_{\gamma \gamma} \times B(\chi_{c0} \to D \bar{D}) 
= 
2.171~{\rm keV} \times 0.328 = 0.71~{\rm keV} \,.
\label{Gamma_gamgam_BR_chic0}
\end{equation}
It should be emphasized that the estimate of 
$\Gamma_{\gamma \gamma} \times B(\chi_{c0} \to D \bar{D})$ 
is highly uncertain in view of the uncertainty of $\Gamma_{\gamma \gamma}$ 
(dependent on the choice of potential model) 
and $B(\chi_{c0} \to D \bar{D})$
regarding the choice of total width
$\Gamma_{\chi_{c0}}$, 
(\ref{4.0_pdg}) or (\ref{4.0_fit}),
and of $\Gamma(\chi_{c0} \to D \bar{D})$
(see the discussion above).
Note that the present experimental widths are not well determined.
Clearly, more investigations are necessary
to confirm the property of the $\chi_{c0}(2P)$.

\subsection{Continuum production of neutral $D$ mesons: $\gamma^{*} \gamma^{*} \to D^{0} \bar{D}^{0}$}
\label{sec:Formalism_B}

The $\gamma \gamma \to D^{0} \bar{D}^{0}$ amplitude
with the $D^{*0} \equiv D^{*}(2007)^{0}$ 
$t/u$-channel exchanges can be expressed as
\begin{eqnarray}
{\cal M}_{\mu \nu}(\gamma^{*} \gamma^{*} \to D^0 \bar D^0)
&=& 
i\Gamma^{(D^{*0}D^0\gamma)}_{\kappa_{1} \mu}(\hat{p}_{t},q_{1})\,
i\tilde{\Delta}^{(D^{*0})\,\kappa_{1} \kappa_{2}}(s_{34},\hat{p}_{t}^{2})\,
i\Gamma^{(D^{*0}D^0\gamma)}_{\kappa_{2} \nu}(-\hat{p}_{t},q_{2}) 
\nonumber \\
&&
+ \,
i\Gamma^{(D^{*0}D^0\gamma)}_{\kappa_{1} \mu}(-\hat{p}_{u},q_{1})\,
i\tilde{\Delta}^{(D^{*0})\,\kappa_{1} \kappa_{2}}(s_{34},\hat{p}_{u}^{2})\,
i\Gamma^{(D^{*0}D^0\gamma)}_{\kappa_{2} \nu}(\hat{p}_{u},q_{2}) \,,
\label{amplitude_D0D0bar_continuum}
\end{eqnarray}
where 
$\hat{p}_{t} = p_{a} - p_{1} - p_{3}$,
$\hat{p}_{u} = p_{4} - p_{a} + p_{1}$, and
$s_{34} = p_{34}^{2}$.

The $D^{*0}D^{0}\gamma$ vertex, including a form factor, 
is taken as:
\begin{eqnarray}
i\Gamma_{\mu \nu}^{(D^{*0}D^0\gamma)}(\hat{p},q) 
= -i e \frac{g_{D^{*0}D^0\gamma}}{m_{D^{*0}}}\, 
\varepsilon_{\mu \nu \rho \sigma} \hat{p}^{\rho} q^{\sigma}
F^{(D^{*0}D^{0}\gamma)}(\hat{p}^{2},q^{2})\,.
\label{DstarDgam_vertex}
\end{eqnarray}
We use the factorized form for the $D^{*0}D^{0}\gamma$ form factor
\begin{eqnarray}
F^{(D^{*0}D^0\gamma)}(\hat{p}^{2},q^{2}) = 
F^{(D^{*0})}(\hat{p}^{2}) F^{(\gamma)}(q^{2}) 
\label{DstarDgam_ff}
\end{eqnarray}
with $F^{(D^{*0}D\gamma)}(m_{D^{*0}}^{2},0) = 1$.
We take 
\begin{eqnarray}
F^{(D^{*0})}(\hat{p}^{2}) &=& \exp \left( \frac{\hat{p}^{2}-m_{D^{*0}}^{2}}{\Lambda_{D^{*0}}^{2}} \right)\,,
\label{DstarDgam_ff_exp}
\\
F^{(\gamma)}(q^{2}) &=& \frac{\Lambda_{\gamma}^{2}}{\Lambda_{\gamma}^{2}-q^{2}}\,,
\qquad \Lambda_{\gamma} = 1\;{\rm GeV}\,.
\label{DstarDgam_ff_aux}
\end{eqnarray}
Here $\hat{p}^{2} < 0$ and $q^{2} < 0$.
We take $\Lambda_{D^{*0}} = 3.3 - 3.5\;{\rm GeV}$
estimated from a comparison of the model results
to the experimental data 
(see Figs.~\ref{fig:M34_BaBar}--\ref{fig:M34_BaBar_eff}).

The coupling constant $g_{D^{*0}D^0\gamma}$ 
is related to the decay width of $D^{*}(2007)^{0} \to D^{0} \gamma$
as follows
\begin{eqnarray}
\Gamma(D^{*0} \to D^{0} \gamma) = 
\frac{\alpha_{\rm em}}{24}\frac{(m_{D^{*0}}^{2}-m_{D^{0}}^{2})^{3}}{m_{D^{*0}}^{5}}
|g_{D^{*0} D^{0} \gamma}|^{2}\,.
\label{coupling_DstarDgam}
\end{eqnarray}
From \cite{ParticleDataGroup:2024cfk} we find
\begin{eqnarray}
&&m_{D^{*0}} = 2006.85 \pm 0.05~{\rm MeV}\,, \nonumber \\
&&m_{D^{0}} = 1864.84 \pm 0.05~{\rm MeV}\,, \nonumber \\
&&\Gamma(D^{*0} \to D^{0} \gamma) / \Gamma_{D^{*0}} 
= (35.3 \pm 0.9)\%\,,  \nonumber \\ 
&&\Gamma_{D^{*0}} < 2.1~{\rm MeV} \,.
\label{PDG_Ds2007}
\end{eqnarray}
There is no accurate experimental result 
for the total width of $D^{*}(2007)^{0}$. 
However, there are some estimates, see e.g.
\cite{Jaus:1996np,Rosner:2013sha,Guo:2019qcn,Jia:2024imm}. 
With $\Gamma_{D^{*0}} = 55.3$~keV \cite{Guo:2019qcn,Cao:2024nxm}
we get $|g_{D^{*0} D^{0} \gamma}| = 5.97$
and this value is used in our calculations.
The sign of $g_{D^{*0} D^{0} \gamma}$
does not matter as our amplitude is quadratic in the coupling.

We use in (\ref{amplitude_D0D0bar_continuum})
$\tilde{\Delta}^{(D^{*0})\, \kappa_{1} \kappa_{2}}(s_{34},\hat{p}^{2})$
the propagators
for the reggeized vector meson $D^{*}(2007)^{0}$.
Two scenarios for the vector-meson reggeization effect were considered in \cite{Lebiedowicz:2019jru,Lebiedowicz:2021pzd}.
In the present work, we proceed analogously to (3.20)--(3.22) of \cite{Lebiedowicz:2019jru}.
The $D^{*0}$-meson propagator and the transverse function reads
\begin{eqnarray}
&&\tilde{\Delta}^{(D^{*0})\, \kappa_{1} \kappa_{2}}(s_{34},\hat{p}^{2})
= - g^{\kappa_{1} \kappa_{2}} \tilde{\Delta}_{T}^{(D^{*0})}(s_{34},\hat{p}^{2})
\,, \\
&&\tilde{\Delta}_{T}^{(D^{*0})}(s_{34},\hat{p}^{2})=
\Delta_{T}^{(D^{*0})}(\hat{p}^{2})
\left( \exp (i \phi(s_{34}))\,\frac{s_{34}}{s_{\rm thr}} 
\right)^{\alpha_{D^{*}}(\hat{p}^{2})-1}\,,
\label{reggeization}
\end{eqnarray}
where we take the simple expression
$(\Delta_{T}^{(D^{*0})}(\hat{p}^{2}))^{-1} = \hat{p}^{2} - m_{D^{*0}}^{2}$, and
\begin{eqnarray}
\phi(s_{34}) =
\frac{\pi}{2}\exp\left(\frac{s_{\rm thr}-s_{34}}{s_{{\rm thr}}}\right)-\frac{\pi}{2}\,, \qquad
s_{\rm thr} = 4 m_{D^{0}}^2\,.
\label{reggeization_aux}
\end{eqnarray}
We use a specific nonlinear Regge trajectory for the $D^{*0}$ mesons,
the so-called ``square-root'' trajectory \cite{Brisudova:1999ut}, viz.
\begin{eqnarray}
\alpha_{D^{*}}(\hat{p}^{2}) = 
\alpha_{D^{*}}(0) + 
\gamma \left( \sqrt{T_{D^{*}}} - \sqrt{T_{D^{*}} - \hat{p}^{2}}  \right)\,,
\label{Kstar_trajectory_nonlinear}
\end{eqnarray}
where $\gamma$ governs the slope of the trajectory
and $T_{D^{*}}$ denotes the trajectory termination point.
The parameters are fixed to be 
$\alpha_{D^{*}}(0) = -1.02$,
$\gamma = 3.65$~GeV$^{-1}$, $\sqrt{T_{D^{*}}} = 3.91$~GeV;
see Table~I of \cite{Brisudova:1999ut}.

\subsection{Continuum production of charged $D$ mesons: $\gamma^{*} \gamma^{*} \to D^{+} D^{-}$}
\label{sec:Formalism_C}

The continuum amplitude with the $D^{\pm}$ exchanges
can be written in a similar manner as described in 
\cite{Lebiedowicz:2015cea,Klusek-Gawenda:2017lgt}
for the $\gamma \gamma \to H^{+} H^{-}$ 
and $\gamma \gamma \to p \bar{p}$ collisions, respectively.
The amplitude for the subprocess $\gamma \gamma \to D^{+} D^{-}$ has the form:
\begin{eqnarray}
{\cal M}_{\mu \nu}(\gamma^{*} \gamma^{*} \to D^{+} D^{-})
&=& -i e^{2}
\left[ 
(q_{1} - 2 p_{3})_{\mu}(q_{1} - p_{3} + p_{4})_{\nu} \frac{1}{\hat{p}_{t}^{2} - m_{D}^{2}}
+
(q_{1} - 2 p_{4})_{\mu}(q_{1} + p_{3} - p_{4})_{\nu} \frac{1}{\hat{p}_{u}^{2} - m_{D}^{2}}
-
2 g_{\mu \nu}
\right] \nonumber \\
&& \times 
F^{(\gamma)}(q_{1}^{2}) F^{(\gamma)}(q_{2}^{2})
F(\hat{p}_{t}^{2},\hat{p}_{u}^{2},p_{34}^{2})\,.
\label{amplitude_DpDm_continuum}
\end{eqnarray}
Here $F^{(\gamma)}(q_{1,2}^{2})$ are given by
(\ref{DstarDgam_ff_aux}) and
the off-shell meson dependences are taken into account via multiplication of the amplitude
by a common form factor
$F(\hat{p}_{t}^{2},\hat{p}_{u}^{2},p_{34}^{2})$
[see (2.12), (2.13) of \cite{Klusek-Gawenda:2017lgt}
but with the replacements $m_{p} \to m_{D^{+}}$ and $\Lambda_{p} \to \Lambda_{D}$].
The parameter $\Lambda_{D}$ should be fitted to the experimental data.
In the present paper, we performed calculations for 
$F(\hat{p}_{t}^{2},\hat{p}_{u}^{2},p_{34}^{2}) = 1$.

By a similar way as in Sec.~\ref{sec:Formalism_B}, 
the $D^{+} D^{-}$-production mechanism via the $D^{*}(2010)^{\pm}$ exchanges
can also be considered.
To estimate the coupling constant
$g_{D^{*}(2010)^{+}D^{+}\gamma}$
we find \cite{ParticleDataGroup:2024cfk}
\begin{eqnarray}
&&m_{D^{*}(2010)^{+}} = 2010.26 \pm 0.05~{\rm MeV}\,, \nonumber \\
&&m_{D^{+}} = 1869.66 \pm 0.05~{\rm MeV}\,, \nonumber \\
&&\Gamma(D^{*}(2010)^{+} \to D^{+} \gamma) / \Gamma_{D^{*0}(2010)^{+}} 
= (1.6 \pm 0.4)\%\,,  \nonumber \\ 
&&\Gamma_{D^{*}(2010)^{+}} = 83.4 \pm 1.8~{\rm keV} \,.
\label{PDG_Ds2010}
\end{eqnarray}
Using (\ref{coupling_DstarDgam}) 
with (\ref{PDG_Ds2010}) we get
$|g_{D^{*}(2010)^{+}D^{+}\gamma}| = 0.3$.
Thus, the corresponding contribution of the continuum
for 
$e^{+} e^{-} \to e^{+} e^{-} D^{+} D^{-}$
via the $D^{*}(2010)^{\pm}$ exchanges
is much smaller than for the $D^{0} \bar{D}^{0}$ case,
and can be safely neglected.

\section{Results and discussion}
\label{sec:Results}

The experimental cross section (efficiency corrected)
for $\chi_{c2}(3930)$ production, 
determined by the BaBar Collaboration 
(Eq.~(17) of \cite{BaBar:2010jfn}), is
\begin{eqnarray}
\sigma_{\rm exp}(e^+ e^- \to e^+ e^- (\gamma \gamma \to \chi_{c2}(3930) \to D \bar{D})) 
= 741 \pm 166 \; {\rm fb}\,.
\label{xs_BaBar}
\end{eqnarray}
Here the error is only statistical.
It is not clear whether the BaBar Collaboration imposed the cut on $p_{t}$ of the $D \bar{D}$ pair when presenting the integrated cross section.

The BaBar~\cite{BaBar:2010jfn} and Belle~\cite{Belle:2005rte}
Collaborations 
conclude that
\begin{equation}
\Gamma_{\gamma \gamma} \times B(\chi_{c2} \to D \bar{D}) 
= 
\begin{cases}
0.24 \pm 0.05~{\rm (stat)} \pm 0.04~{\rm (syst)} \, {\rm keV} & {\rm from~BaBar} \,,
  \\
0.18 \pm 0.05~{\rm (stat)} \pm 0.03~{\rm (syst)} \, {\rm keV} & {\rm from~Belle}  \,,
 \end{cases}
\label{Gamma_gamgam}
\end{equation}
with the assumption for the Belle result that
$B(\chi_{c2} \to D^{+} D^{-}) = 0.89 \times B(\chi_{c2} \to D^{0} \bar{D}^{0})$.
The PDG average gives
$\Gamma_{\gamma \gamma} \times B(\chi_{c2} \to D \bar{D}) = 0.21 \pm 0.04$~keV \cite{ParticleDataGroup:2024cfk}.

\subsection{$e^+ e^- \to e^+ e^- \chi_{c2}(3930)$}
\label{sec:3b}

We made calculations of the cross sections for the $e^+ e^- \to e^+ e^- \chi_{c2}(3930)$ reaction for c.m. energy $\sqrt{s} = 10.54$~GeV.
The results are 
$\sigma(e^+ e^- \to e^+ e^- \chi_{c2}) = 1.577$~pb
using the Cornell potential,
and $1.022$~pb using Buchm\"uller-Tye (BT) potential.
To obtain (\ref{xs_BaBar}), 
we find $B(\chi_{c2} \to D \bar{D})$
corresponding to $\Gamma_{\gamma \gamma}$
from Table~\ref{tab:R0_LFWF} for these two potentials.
We got 
\begin{equation}
\Gamma_{\gamma \gamma} \times B(\chi_{c2} \to D \bar{D}) 
= 
\begin{cases}
0.840~{\rm keV} \times 0.47^{+0.10}_{-0.11} = 0.39^{+0.09}_{-0.09}~{\rm keV} 
& {\rm for~Cornell} \,,
\\
0.544~{\rm keV} \times 0.73^{+0.16}_{-0.17} = 0.40^{+0.08}_{-0.10}~{\rm keV} 
& {\rm for~BT} \,.
\end{cases}
\label{Gamma_gamgam_th}
\end{equation}
Comparing the results (\ref{Gamma_gamgam_th})
to the BaBar result (\ref{Gamma_gamgam})
($\Gamma_{\gamma \gamma} \times B(\chi_{c2}(2P) \to D \bar{D}) = 0.24 \pm 0.06~{\rm keV}$)
one can see that they are roughly consistent within the uncertainty.

Taking into account the condition
$p_{t, \chi_{c2}} < 0.05$~GeV
we obtain 
$\sigma(e^+ e^- \to e^+ e^- \chi_{c2}(2P)) = 0.625$~pb
(Cornell)
and $0.405$~pb (BT).
These results with comparison to (\ref{Gamma_gamgam})
leads to $B(\chi_{c2} \to D \bar{D}) = 1.19$ and 1.83
for Cornell and BT potential, respectively.
This suggests that the value given in (\ref{xs_BaBar}) does not include cuts.


\subsection{$e^+ e^- \to e^+ e^- D \bar{D}$}
\label{sec:3c}

In Table~\ref{tab:2}, we listed the values of the fitted parameter
$|g_{\chi_{c2} D \bar{D}}|$ obtained from comparison to
the combined BaBar cross section (\ref{xs_BaBar}).
For neutral and charged channel we obtain
\begin{align}
\sigma(e^+ e^- \to e^+ e^- (\chi_{c2}(3930) \to D^{0} \bar{D}^{0})) 
&= 390 \pm 87 \;{\rm fb}\,, \nonumber \\
\sigma(e^+ e^- \to e^+ e^- (\chi_{c2}(3930) \to D^{+} D^{-})) 
&= 351 \pm 79 \;{\rm fb}\,.
\end{align}
The requirement of very small transverse momentum 
$p_{t, D\bar{D}} < 0.05$~GeV
is motivated by an untagged analysis 
of $\gamma \gamma \to D \bar{D}$ production by Belle and BaBar.
Due to the $p_{t, D\bar{D}}$ cut the cross section
is reduced by a factor of 2.54.
In order to describe the BaBar result (\ref{xs_BaBar}), 
this causes a larger value for the constant $|g_{\chi_{c2} D \bar{D}}|$
and thus leads to branching fraction $B(\chi_{c2} \to D \bar{D}) \sim 1$.
Then, the theoretical results for the potential models 
under consideration are not consistent with the experiment.
\begin{table}[h!]
\caption{Results for the reaction
$e^+ e^- \to e^+ e^- D \bar{D}$ via $\chi_{c2}(3930)$ 
calculated for $\sqrt{s} = 10.54$~GeV
without and with the limitations on the transverse momenta of the final state particles.
The fitted coupling parameter $|g_{\chi_{c2} D \bar{D}}|$
for the Cornell and Buchm\"uller-Tye (BT) potential is shown.
In the last column the product of the two-photon width
of the $\chi_{c2}(3930)$ state
and the branching fraction to $D \bar{D}$
for each model is given.}
    \centering
    \begin{tabular}{l |l |r |c |c }
    \hline
    \hline
     &  
     Potential &
     $|g_{\chi_{c2} D \bar{D}}|$ &
     $B(\chi_{c2} \to D \bar{D})$&
     $\Gamma_{\gamma \gamma} \times B(\chi_{c2} \to D \bar{D})$ (keV) \\
     \hline
no cuts   
& Cornell 
& $8.1^{+0.9}_{-1.0}$ 
& $0.38 \pm  0.08$ & $0.32 \pm  0.07$\\
& BT                  
& $10.1^{+1.1}_{-1.2}$ 
& $0.58 \pm  0.13$ & \\
     \hline
$p_{t1}, p_{t2} < 1$~GeV    
& Cornell             
& $8.4^{+0.9}_{-1.0}$ 
& $0.40 \pm  0.09$ & $0.34 \pm  0.08$\\
& BT                  
& $10.4^{+1.1}_{-1.2}$ 
& $0.62 \pm  0.14$ &\\
     \hline
$p_{t, D\bar{D}} < 0.05$~GeV    
& Cornell             
& $12.9^{+1.4}_{-1.5}$ 
& $0.96 \pm  0.21$ & $0.80 \pm  0.18$\\
& BT                  
& $16.1^{+1.7}_{-1.9}$ 
& $1.48 \pm  0.33$ & \\
         \hline
         \hline
    \end{tabular}
    \label{tab:2}
\end{table}

From Table~\ref{tab:2}
one can see that the obtained value of the branching fraction 
$B(\chi_{c2}(2P) \to D \bar{D})$ depends on
the potential model 
(the two-photon width of the $\chi_{c2}(3930)$ state)
assumed in the calculations.
For the results calculated without kinematical cuts, 
the obtained value for the product
$\Gamma_{\gamma \gamma} \times B(\chi_{c2}(2P) \to D \bar{D})$
is consistent with existing Belle and BaBar measurements within uncertainties.
From our analysis, the corresponding branching fraction is
$B(\chi_{c2}(3930) \to D \bar{D}) = 0.38 \pm 0.08$ 
for the Cornell potential and 
$0.58 \pm 0.13$ for the Buchm\"uller-Tye potential.
The latter value is compatible 
with the one reported in \cite{Swanson:2006st}.

\begin{figure}
\includegraphics[width=0.325\textwidth]{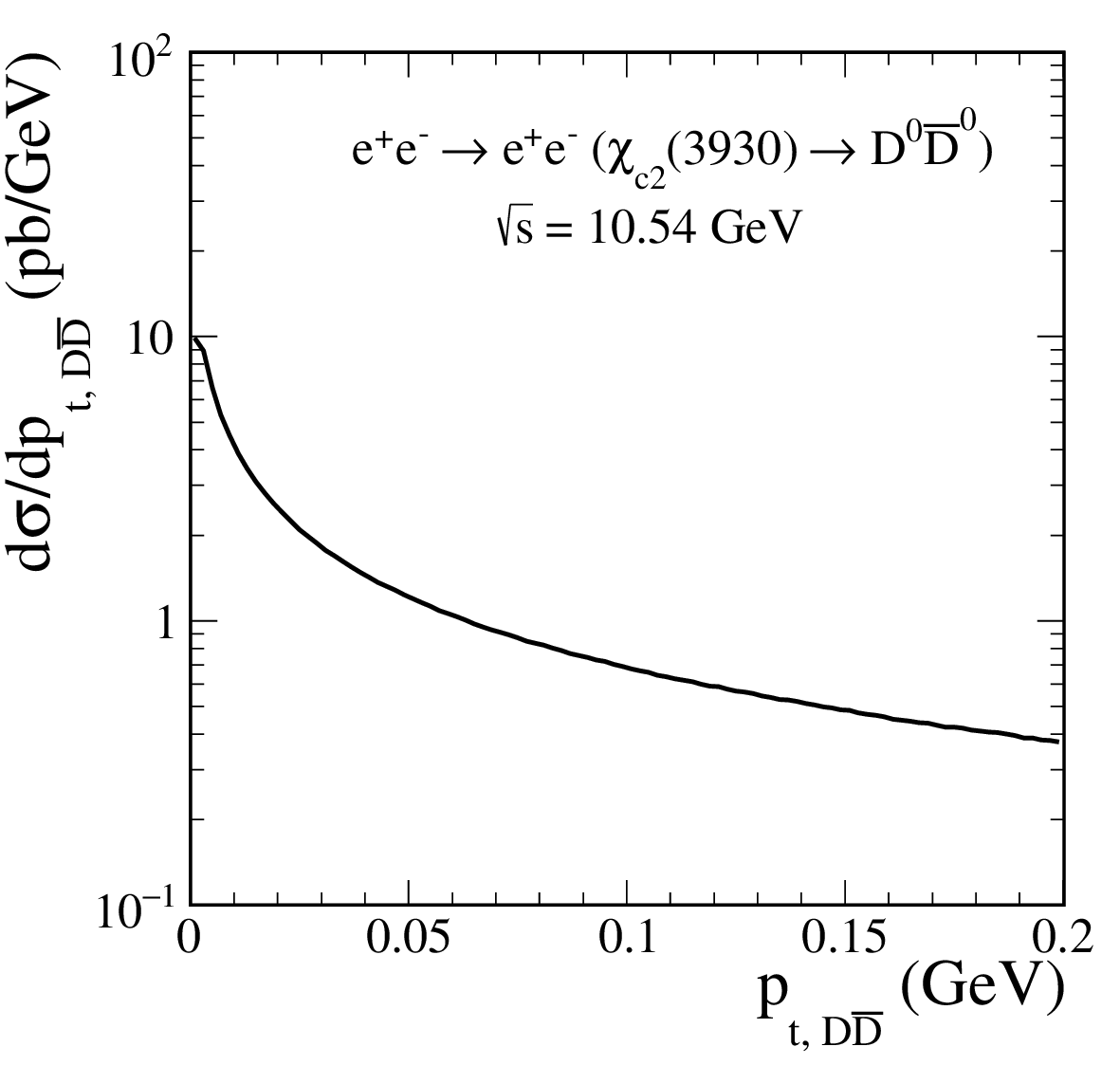}
\includegraphics[width=0.325\textwidth]{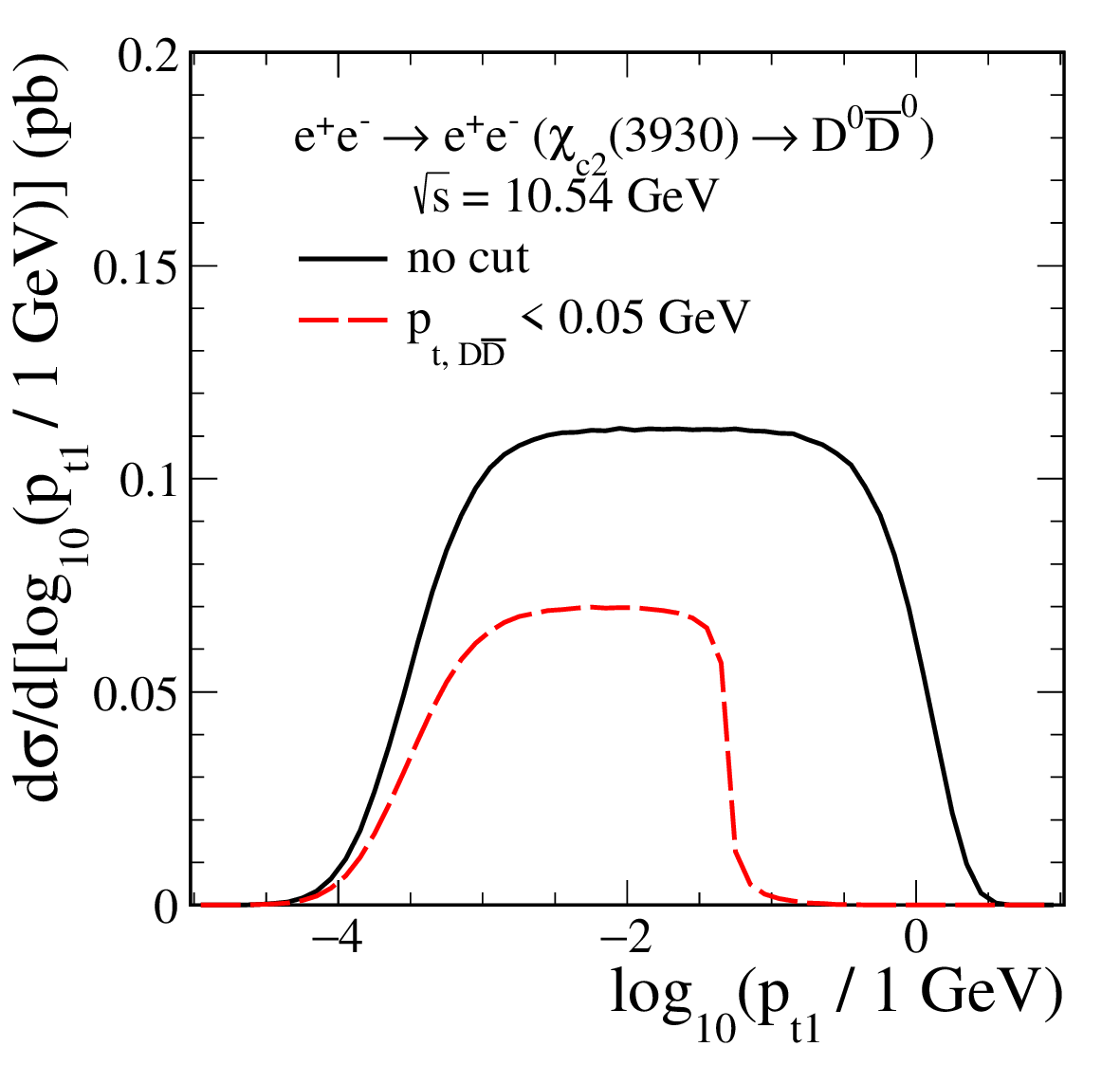}
\includegraphics[width=0.325\textwidth]{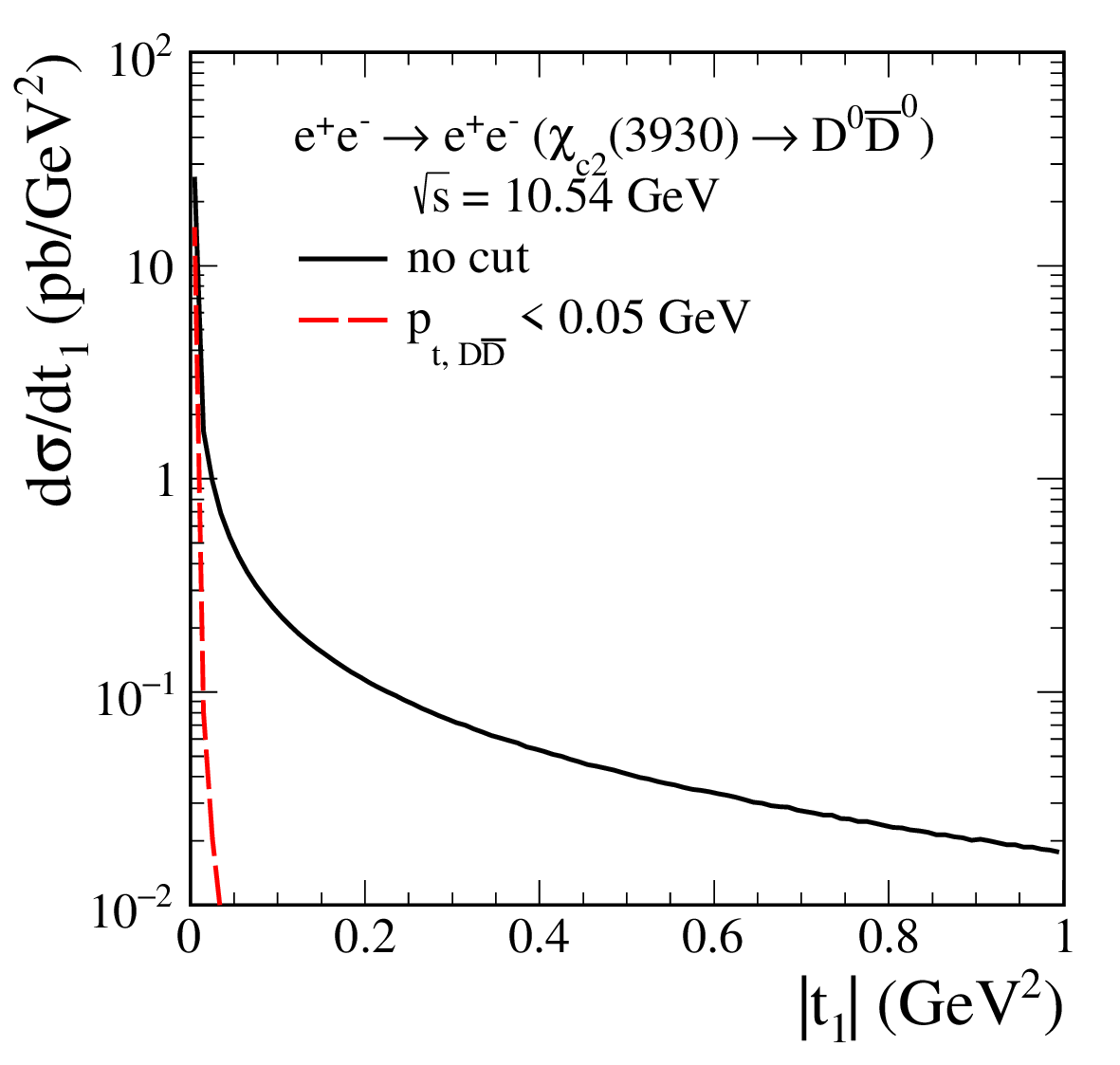}
\caption{Distributions in $p_{t, D\bar{D}}$,
$\log_{10}(p_{t1}/ 1\,\rm{GeV})$ and
$|t_{1}|$ 
for the $e^+ e^- \to e^+ e^- (\chi_{c2}(3930) \to D^0 {\bar D}^0)$ reaction
for $\sqrt{s} = 10.54$~GeV.
We show the results without and with the cut on $p_{t, {\rm sum}} < 0.05$~GeV.
The calculations were done for the BT potential and with
$g_{\chi_{c2} D \bar{D}} = 10.1$.}
\label{fig:dsig_dxi}
\end{figure}

Now we wish to show some differential distributions
for the $e^+ e^- \to e^+ e^- (\chi_{c2}(3930) \to D^0 {\bar D}^0)$ reaction.
Figure~\ref{fig:dsig_dxi} shows the distributions in
$p_{t, D\bar{D}}$ the transverse momentum of the $D^0 {\bar D}^0$ pair,
in $\log_{10}(p_{t1}/ 1\,\rm{GeV})$
one of the integration variables, and
in $|t_{1}|$ the four-momentum transfer squared.
In the calculation
we used $g_{\chi_{c2} D \bar{D}} = 10.1$
corresponding to the BT potential.
We show results when
we have not imposed any restrictions on kinematic variables
and the results with the condition
$p_{t, D\bar{D}} < 0.05$~GeV.
The distribution in $p_{t, D\bar{D}}$
peaks in the region of low $p_{t, D\bar{D}}$.
The $t$~distributions are strongly peaked 
at very small $|t_{1}|$ and $|t_{2}|$.
This is caused by the factors $1/t_{1,2}$ from the photon propagators.
One can clearly see that imposing 
the condition on $p_{t, D\bar{D}} < 0.05$~GeV
significantly reduces the cross-section
and limits the momentum transfers to the electrons,
$|t_{1,2}| < 0.01$~GeV$^{2}$.

\begin{figure}
(a)\includegraphics[width=0.4\textwidth]{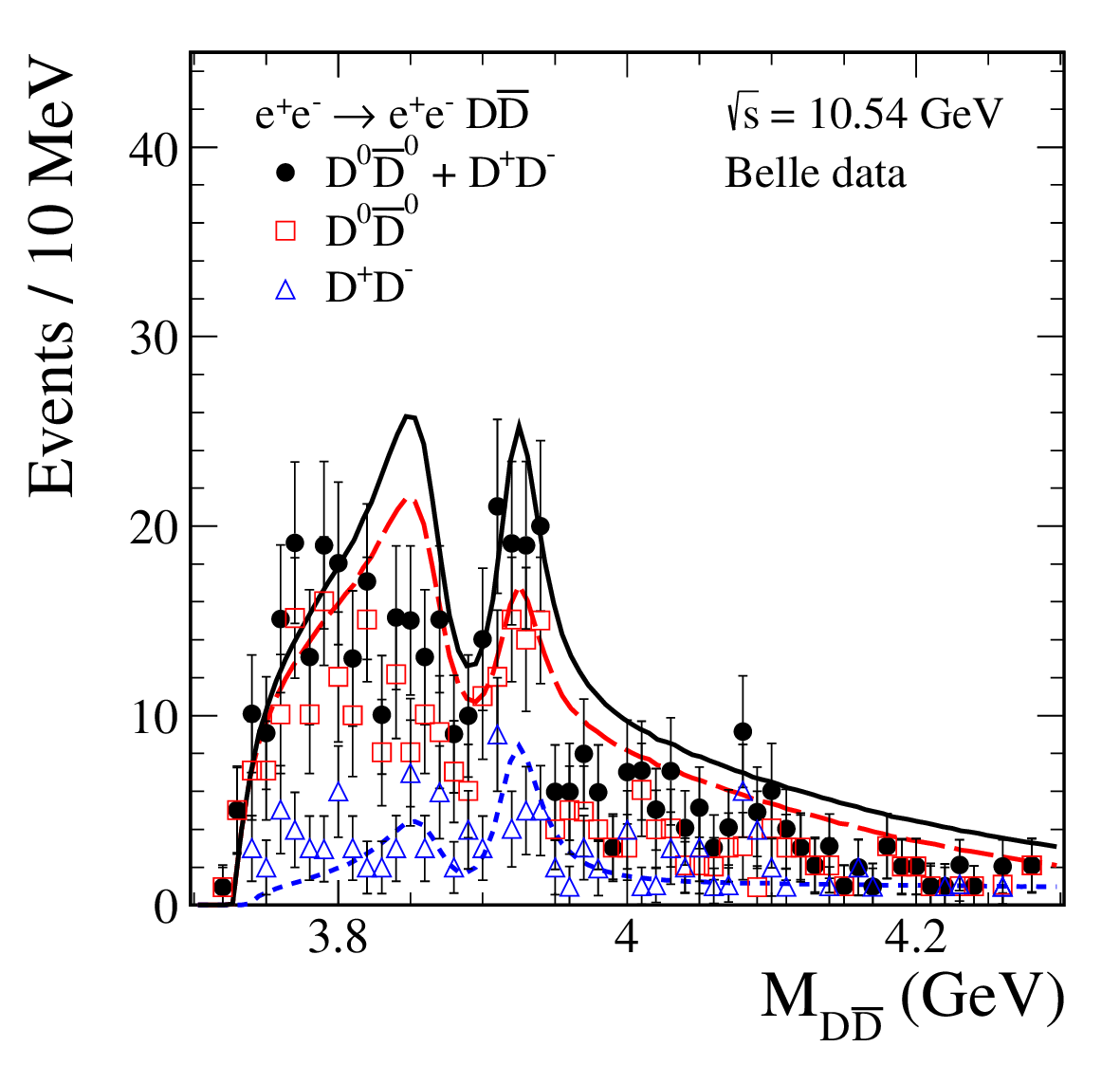}
(b)\includegraphics[width=0.4\textwidth]{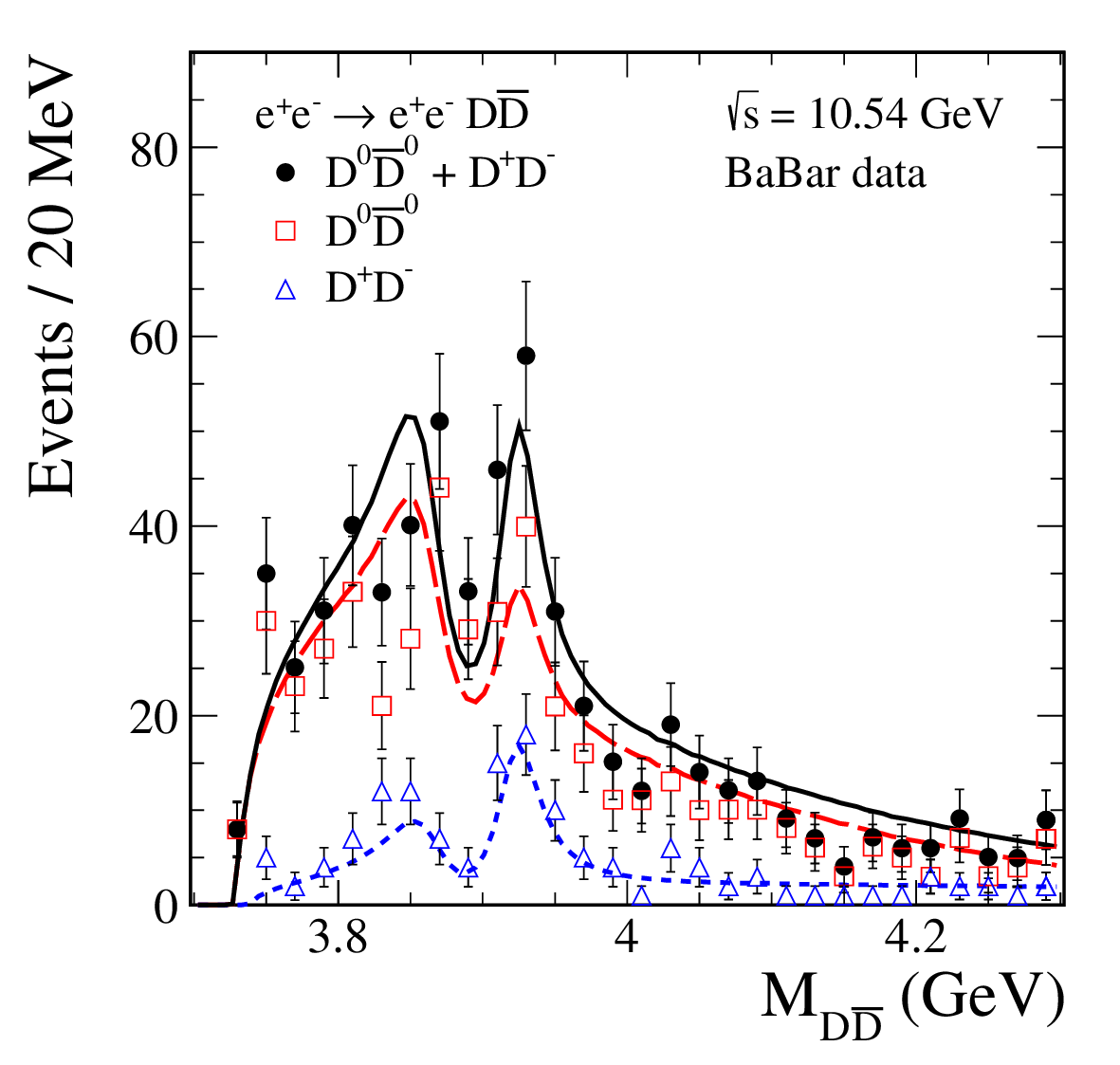}
(c)\includegraphics[width=0.4\textwidth]{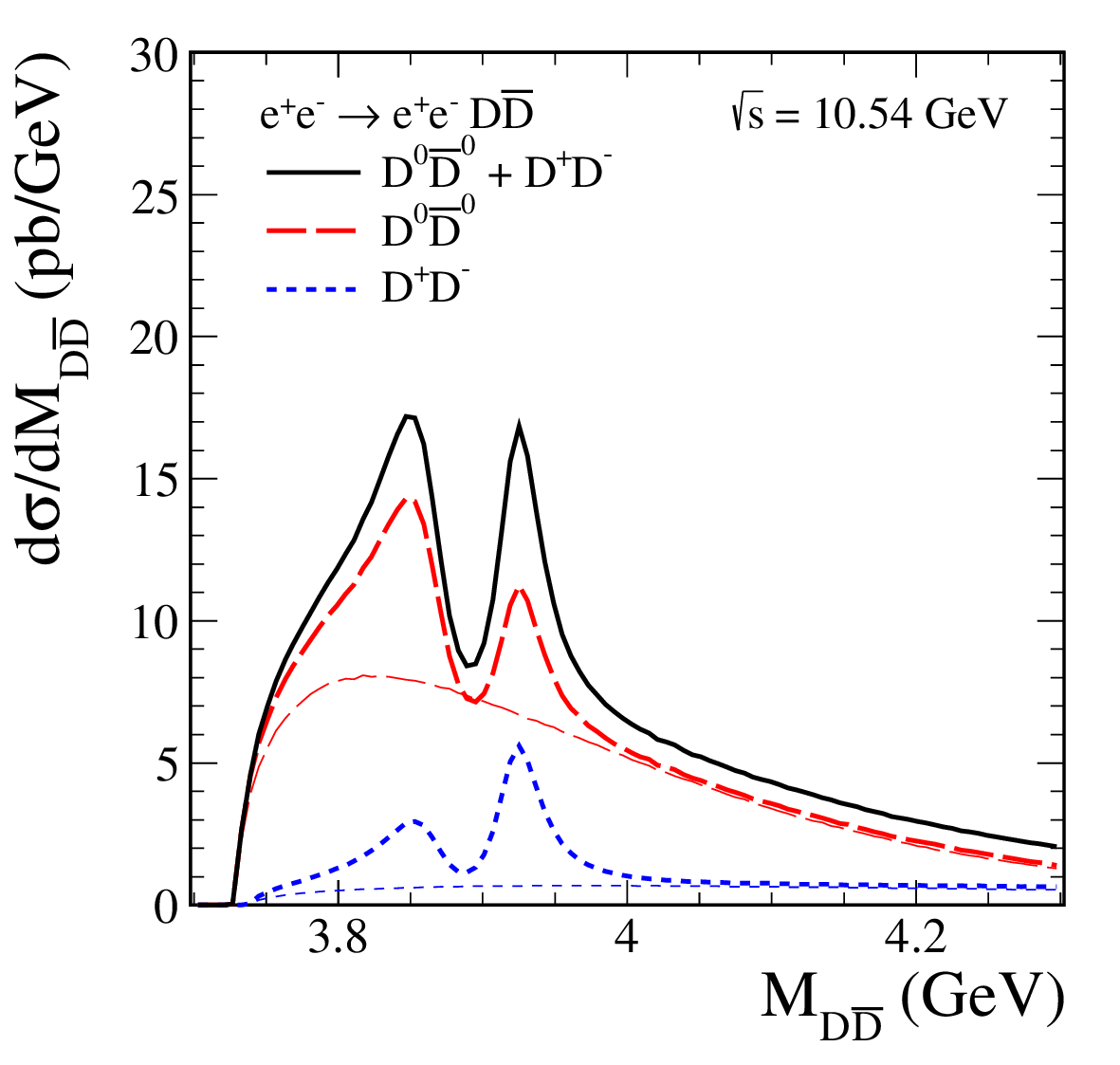}
(d)\includegraphics[width=0.4\textwidth]{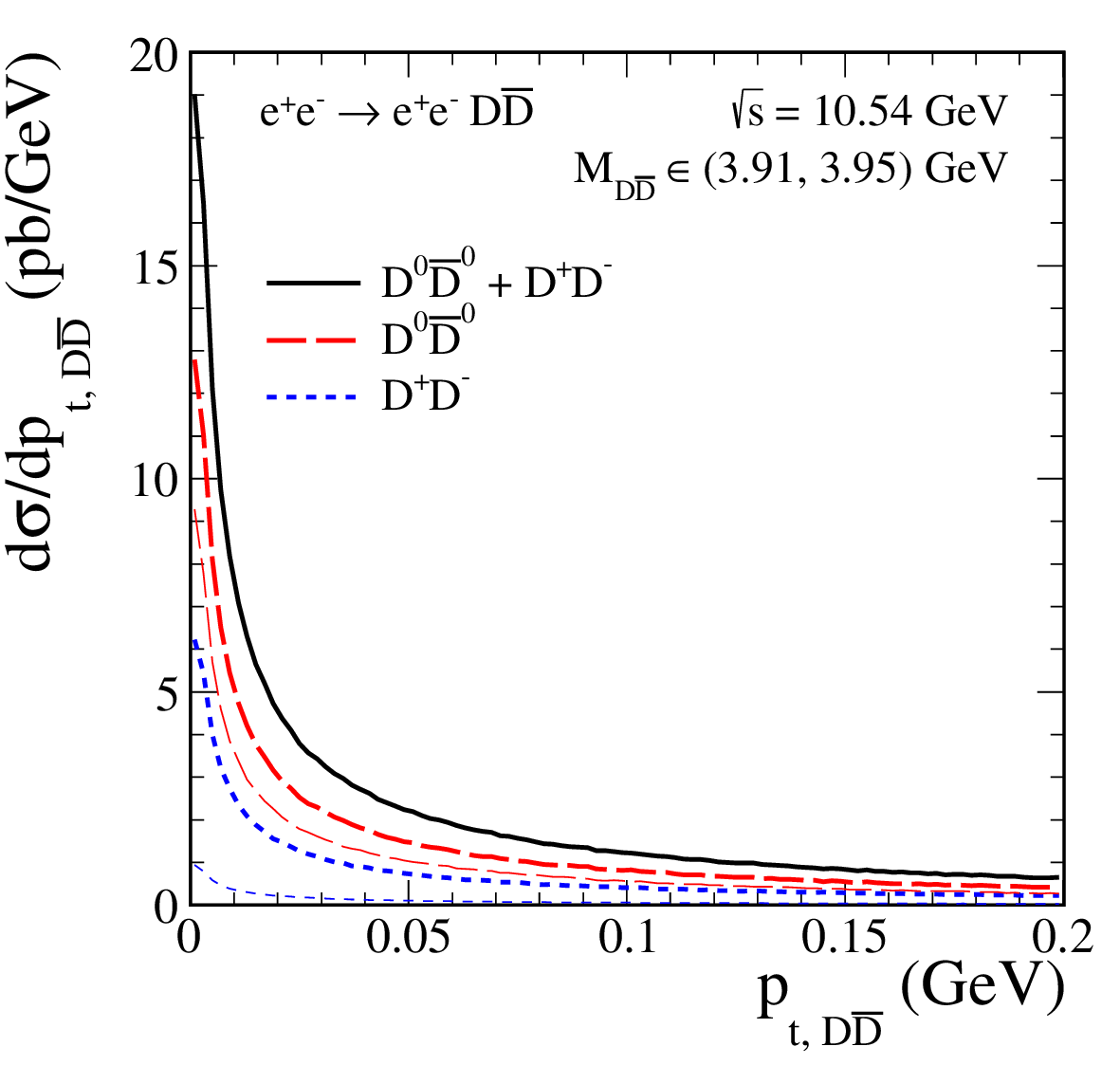}
\caption{
The invariant mass distribution 
for the $e^{+} e^{-} \to e^{+} e^{-} D \bar{D}$ reaction
for the neutral and charged channels,
and the combined $D \bar{D}$ final state.
(a) 
Comparison of theoretical results 
to the Belle data \cite{Belle:2005rte}.
(b) 
Comparison of theoretical results 
to the BaBar data \cite{BaBar:2010jfn}.
(c)
The differential cross sections $d\sigma/dM_{D \bar{D}}$
for the complete model (continuum, $\chi_{c0}(3860)$, and $\chi_{c2}(3930)$ contributions)
and for the continuum contribution alone.
For the $\chi_{c0}(3860)$ term we take (\ref{4.0_fit}).
The red long-dashed lines correspond to the $D^0 {\bar D}^0$ channel
and the blue dashed lines correspond to the $D^{+}D^{-}$ channel.
The black solid line represents the sum of all contributions for two channels.
In panel (d), we show the differential distributions in transverse momentum
of the $D \bar{D}$ pair.
The calculations were done for $\sqrt{s} = 10.54$~GeV
and with the cut on $M_{D \bar{D}} \in (3.91, 3.95)$~GeV.
The meaning of the lines is the same 
as in Fig.~\ref{fig:M34_BaBar}(c).}
\label{fig:M34_BaBar}
\end{figure}

We compare our theoretical results 
for the $e^{+} e^{-} \to e^{+} e^{-} D \bar{D}$ reaction
calculated for $\sqrt{s} = 10.54$~GeV
with the Belle data from Fig.~2 of \cite{Belle:2005rte}
and
BaBar data from Fig.~5 of \cite{BaBar:2010jfn}.
The $\chi_{c0}(3860)$ and $\chi_{c2}(3930)$
contributions are calculated for the BT potential 
with $R'_{2P}(0) = 0.326$~GeV$^{5/2}$
(see Tables~\ref{tab:R0_LFWF} and \ref{tab:chic0_2P})
and with $g_{\chi_{c2} D \bar{D}} = 10.1$
(see Table~\ref{tab:2}),
and $g_{\chi_{c0} D \bar{D}} = 2.52$.
In Fig.~\ref{fig:M34_BaBar}(a)-(c)
we show the invariant mass distributions
for the combined $D \bar{D}$ final state
and for the $D^{0} \bar{D}^{0}$ and $D^{+} D^{-}$ channels separately. 
We can see that the complete model results, 
containing the continuum and resonant contributions, 
can reasonably well describe the experimental data.
There is a significant interference effect between 
the $\chi_{c0}(3860)$ resonance and continuum terms.
The $D^{0} \bar{D}^{0}$-continuum contribution
increases from the reaction threshold, 
reaches a maximum at $M_{D^{0} \bar{D}^{0}} \sim 3.8$~GeV, 
and has a large share in the $\chi_{c2}(3930)$ resonance region.
In these calculations, 
for the $D^{0} \bar{D}^{0}$-continuum contribution, 
we use $\Lambda_{D^{*0}} = 3.5~{\rm GeV}$ in (\ref{DstarDgam_ff_exp}).
From the presented results,
it is clearly visible that 
the $D^{+} D^{-}$-continuum contribution is small. 
Since the form factor 
$F(\hat{p}_{t}^{2}, \hat{p}_{u}^{2}, p_{34}^{2})$
is set to 1 in the calculation, 
our estimate should be viewed as an upper limit, 
especially at large $M_{D^{+} D^{-}}$.
In Fig.~\ref{fig:M34_BaBar}(d) we show the distribution 
in $p_{t, D\bar{D}}$ 
in the region $M_{D \bar{D}} \in (3.91, 3.95)$~GeV.

\begin{figure}
\includegraphics[width=0.4\textwidth]{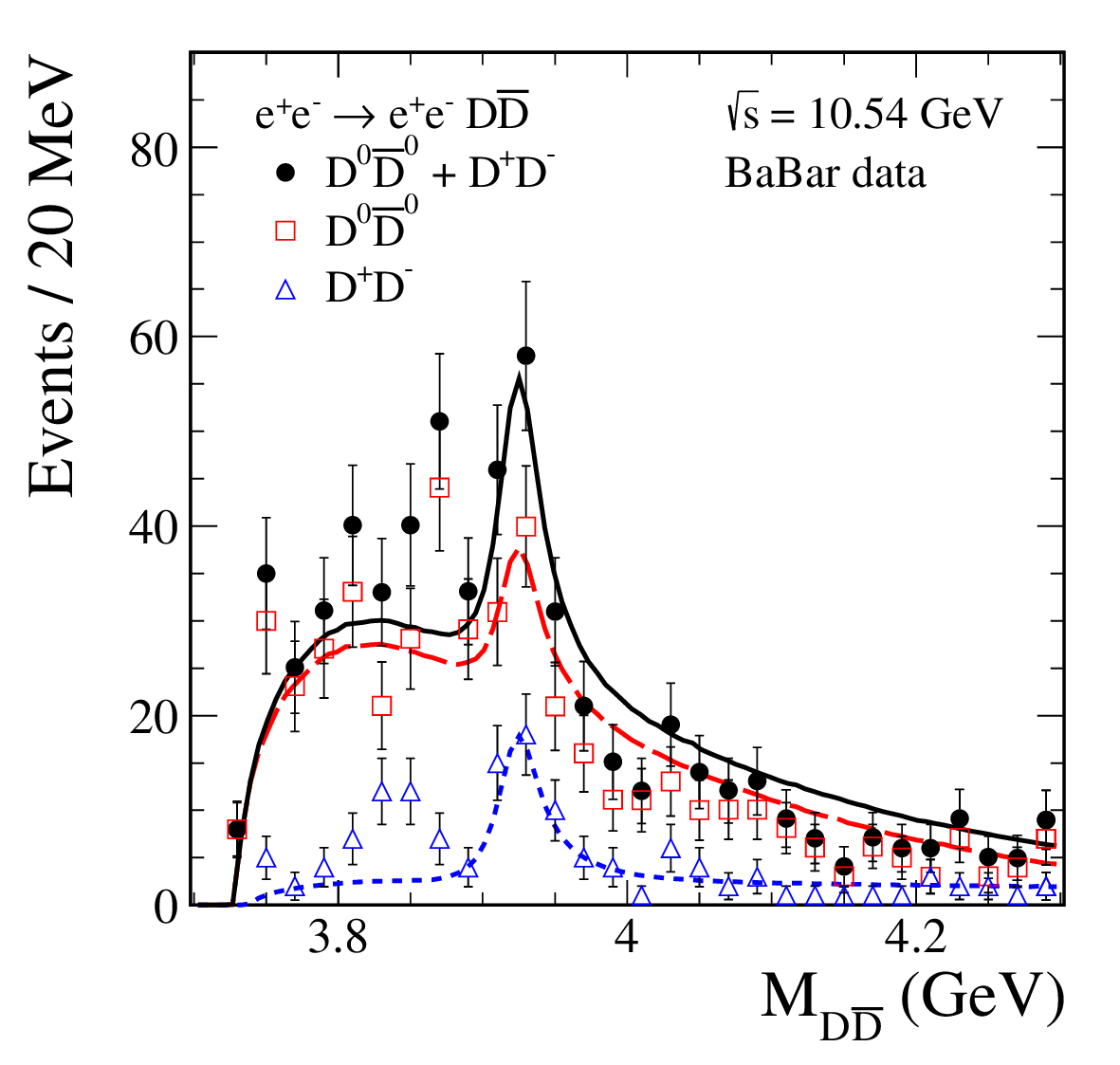}
\includegraphics[width=0.4\textwidth]{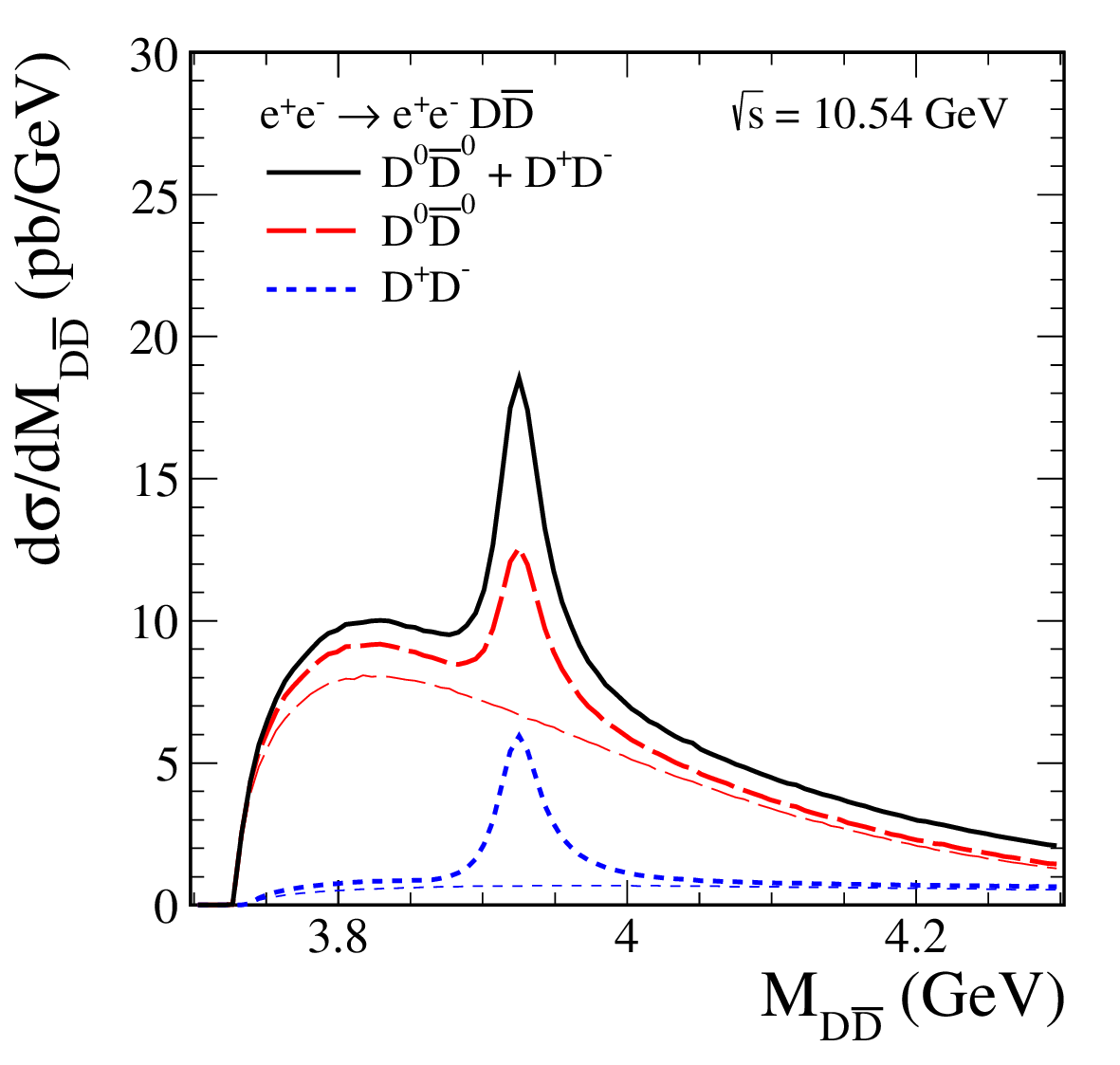}
\caption{
Comparison of theoretical results to the BaBar data \cite{BaBar:2010jfn}.
The meaning of the lines is as in Fig.~\ref{fig:M34_BaBar},
but for the $\chi_{c0}(3860)$ term 
we used the middle values from (\ref{4.0_pdg}).}
\label{fig:M34_BaBar_200}
\end{figure}

For illustrative purposes,
in Fig.~\ref{fig:M34_BaBar_200} we present our results 
obtained for 
$\Gamma_{\chi_{c0}} = 200$~MeV 
(\ref{4.0_pdg}) for the $\chi_{c0}(3860)$ resonance.
In this case, we have that the increase 
in $\Gamma_{\chi_{c0}}$ implies a worse description of low-$M_{D \bar{D}}$ data 
(especially for the $D^{+}D^{-}$ BaBar data).
Here, we keep the coupling constants 
the same as in the previous scenario. 
In general, using the larger total width, 
the coupling constant of $\chi_{c0} \to D \bar{D}$ could be consistently re-estimated. 
If a larger value for $B(\chi_{c0} \to D\bar{D})$ is used, 
this can be compensated for by a smaller two-photon width (see Table~\ref{tab:chic0_2P}).
From the right panel of Fig.~\ref{fig:M34_BaBar_200}
we see that the contributions from the $\chi_{c2}(3930)$ resonance
and $D^{0} \bar{D}^{0}$-continuum are dominant.
The term from $\chi_{c0}(3860)$ is found to be small
but the interference effect between the $\chi_{c0}(3860)$
and the $D^{0} \bar{D}^{0}$ continuum is significant.

An additional comment is in order here.
We are not sure whether the Belle data shown in Fig.~\ref{fig:M34_BaBar}
was corrected for experimental efficiencies. 
If not, this could partly explain why our results are higher than the Belle data. 
Therefore, we have also compared our model results
to the efficiency-corrected BaBar data, 
which was provided only for the combined $D \bar{D}$ final state.

\begin{figure}
\includegraphics[width=0.45\textwidth]{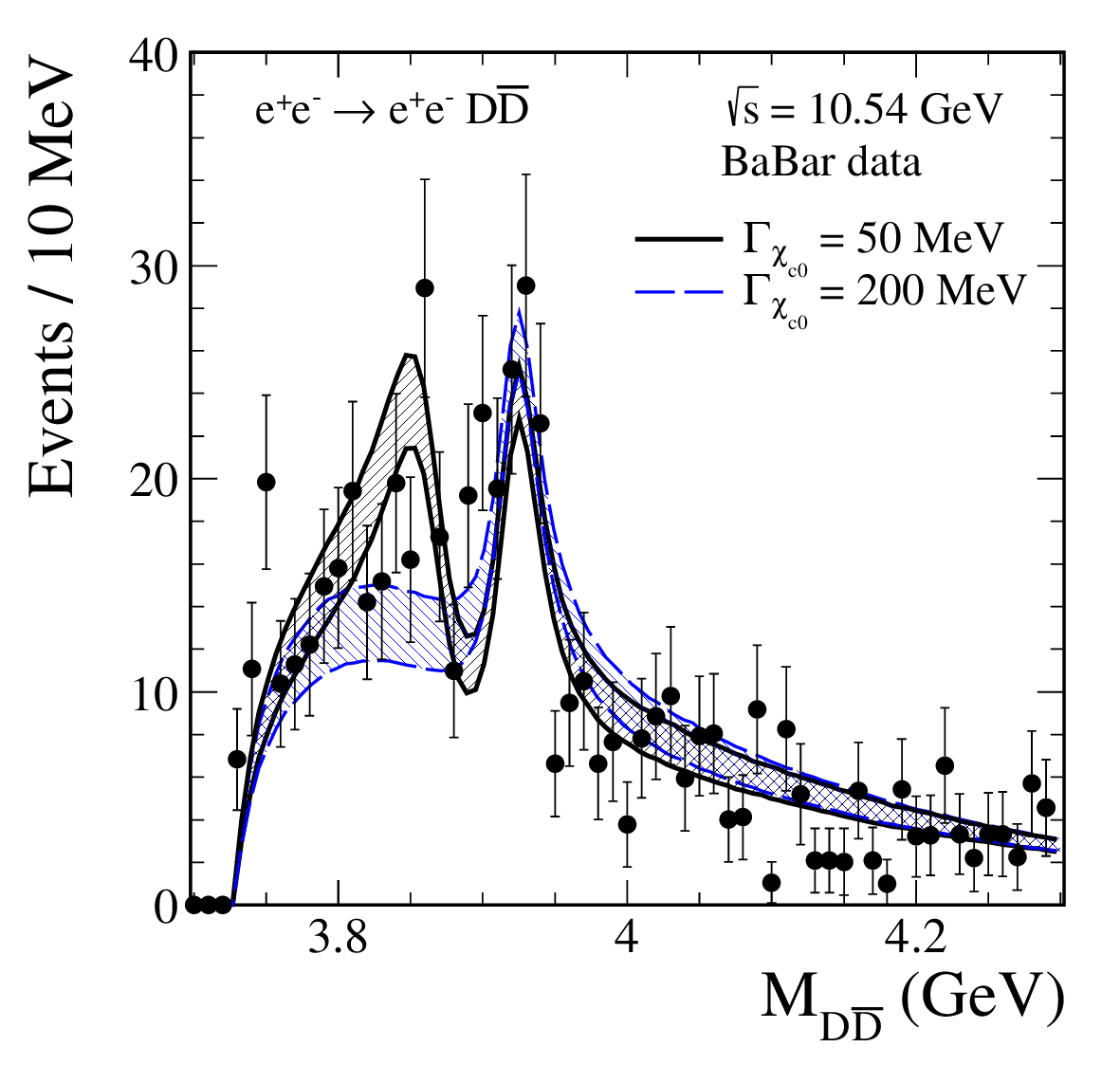}
\caption{
Comparison of our complete model results to the efficiency-corrected 
(combined) BaBar data from Fig.~10 of \cite{BaBar:2010jfn}.
The black solid lines represent results
obtained for $\Gamma_{\chi_{c0}} = 50$~MeV (\ref{4.0_fit}),
while the blue dashed lines correspond to $\Gamma_{\chi_{c0}} = 200$~MeV (\ref{4.0_pdg}).
Here, the calculations for the $D^{0} \bar{D}^{0}$-continuum contribution were made for 
$\Lambda_{D^{*0}} = 3.3$ and $3.5~{\rm GeV}$ in (\ref{DstarDgam_ff_exp}), which correspond to the lower and upper lines in the bands, respectively.
We used the Buchm\"uller-Tye potential
for the excited $2P$ states, $\chi_{c0,2}(2P)$.
}
\label{fig:M34_BaBar_eff}
\end{figure}

Finally, in Fig.~\ref{fig:M34_BaBar_eff} we show our complete model results
including $D^{0} \bar{D}^{0}$ continuum,
$D^{+} D^{-}$ continuum, $\chi_{c0}(3860)$, and $\chi_{c2}(3930)$ contributions
together with the efficiency-corrected (combined) BaBar data 
taken from Fig.~10 of \cite{BaBar:2010jfn}.
In the calculation we used the Buchm\"uller-Tye potential
for the excited $2P$ states, $\chi_{c0,2}(2P)$.
The $D^{0} \bar{D}^{0}$-continuum contribution from 
$t/u$-channel vector-meson $D^{*}(2007)^{0}$ exchanges
depends on the parameter of the off-shell form-factor 
(\ref{DstarDgam_ff_exp}).
In the calculations we take two values of the form-factor parameter
$\Lambda_{D^{*0}} = 3.3~{\rm GeV}$ 
and $3.5~{\rm GeV}$ 
corresponding to the lower and upper line in the bands shown
in Fig.~\ref{fig:M34_BaBar_eff}, respectively.
The presented calculations correspond to two scenarios 
of the $\chi_{c0}(3860)$ total width:
(1) $\Gamma_{\chi_{c0}} = 50$~MeV  
with $\chi^{2}/{\rm dof} = 2.44$ (solid upper line)
and 
$\chi^{2}/{\rm dof} = 1.61$ (solid lower line);
(2) $\Gamma_{\chi_{c0}} = 200$~MeV 
with $\chi^{2}/{\rm dof} = 2.61$ (dashed upper line)
and 
$\chi^{2}/{\rm dof} = 2.06$ (dashed lower line).
We also checked that limiting to the range of
$2 m_{D} \leq M_{D \bar{D}} < 3.9~{\rm GeV}$
the resulting values of $\chi^{2}/{\rm dof}$ are
1.82--2.04 and 1.39--2.85 for scenarios (1) and (2), respectively.

\section{Conclusions}
\label{sec:Conclusions}

In the present paper, we have concentrated on the possible 
observation of excited $P$-wave charmonia 
in the $e^+ e^- \to e^+ e^- D \bar{D}$ reaction.
One of our goals was to estimate the cross-section of meson-pair continua for
both $D^0 {\bar D}^0$ and $D^+ D^-$ channels separately.
We have considered $t$- and $u$-channel exchanges of the vector 
$D^{*0}$ and $D^{*\pm}$ mesons. The corresponding coupling constants
for the $D^* D \gamma$ couplings have been obtained from the experimental 
radiative partial decay widths of $D^* \to D \gamma$.
For the $D^+ D^-$ channel, we have also considered the continuum mechanism
via the $D^{\pm}$ $t/u$-channel exchanges plus contact term.
Comparing our model results with the experimental data 
(Belle \cite{Belle:2005rte}, BaBar \cite{BaBar:2010jfn})
we find a large contribution of the continuum
to the $D^0 {\bar D}^0$ channel and a much smaller one for the $D^+ D^-$ channel
(see Fig.~\ref{fig:M34_BaBar}).

Our analysis suggests that the broad bump at $M_{D \bar D} = 3.8$~GeV  observed in the $D^0 {\bar D}^0$ channel, and not observed significantly in $D^+ D^-$, is rather due to the continuum mechanism.
However, a certain contribution of a broad resonance $\chi_{c0}(3860)$
cannot be completely ruled out.
We have shown that taking this resonance 
into account 
with $\Gamma_{\chi_{c0}} = 50$~MeV
improves 
the description of the near-threshold data,
in particular, the BaBar data 
for the $D^{+}D^{-}$ channel shown in Fig.~\ref{fig:M34_BaBar}(b).
It is rather difficult to understand a large difference between
$\Gamma_{\chi_{c0}}$ and $\Gamma(\chi_{c0} \to D \bar{D})$
in the scenario with large $\Gamma_{\chi_{c0}}$ given in (\ref{4.0_pdg})
which leads to surprisingly small branching fraction
$B(\chi_{c0} \to D \bar{D})$.
It should be noted that the potential-model predictions
for the mass and width of the $\chi_{c0}(2P)$ state differ from the PDG averages.
This indicates that further theoretical and experimental research 
is needed to understand properties of the $\chi_{c0}(2P)$ state more precisely.

We have also estimated the cross section for the $\chi_{c2}(3930)$
production in no tag conditions using $R_{2P}'(0)$ 
calculated within the light front quark model.
Calculations were performed for different potentials
known from the literature; see Table~\ref{tab:R0_LFWF}.
We have obtained an integrated cross-section 
$\sigma(e^+ e^- \to e^+ e^- \chi_{c2}(3930)) \sim 1$~pb
using the Buchm\"uller-Tye potential.
From a comparison of our results 
for the $e^+ e^- \to e^+ e^- (\chi_{c2}(3930) \to D \bar{D})$ reaction
to the BaBar results, we have found
the branching fraction $B(\chi_{c2}(3930) \to D \bar{D}) = 0.58 \pm 0.13$
(Buchm\"uller-Tye potential) and $0.38 \pm 0.08$ (Cornell potential).
We have also obtained
$\Gamma_{\gamma \gamma} \times B(\chi_{c2} \to D \bar{D}) = 0.32 \pm 0.07$~keV 
(see Table~\ref{tab:2}).
This is not far from the experimental value (\ref{Gamma_gamgam}).

Our results strongly motivate a more detailed analysis of the $D \bar{D}$ production.
This would require Monte Carlo simulations of 
both neutral and charged channels, including experimental acceptances and efficiencies.
Hopefully, this will be possible with future Belle~II high-statistics data.




\begin{thebibliography}{99}

\bibitem{ParticleDataGroup:2024cfk}
S.~Navas {\em et~al.}, (Particle Data Group), {\em {Review of particle
  physics},} \href{http://dx.doi.org/10.1103/PhysRevD.110.030001}{Phys. Rev. D
  {\bfseries 110} no.~3, (2024) 030001}.

\bibitem{Cisek:2022uqx}
A.~Cisek, W.~Sch{\"a}fer, and A.~Szczurek, {\em {Structure and production
  mechanism of the enigmatic $X(3872)$ in high-energy hadronic reactions},}
  \href{http://dx.doi.org/10.1140/epjc/s10052-022-11029-x}{Eur. Phys. J. C
  {\bfseries 82} no.~11, (2022) 1062},
  \href{http://arxiv.org/abs/2203.07827}{{arXiv:2203.07827 [hep-ph]}}.

\bibitem{Babiarz:2023ebe}
I.~Babiarz, R.~Pasechnik, W.~Sch{\"a}fer, and A.~Szczurek, {\em {Probing the
  structure of $\chi_{c1}(3872)$ with photon transition form factors},}
  \href{http://dx.doi.org/10.1103/PhysRevD.107.L071503}{Phys. Rev. D {\bfseries
  107} no.~7, (2023) L071503},
  \href{http://arxiv.org/abs/2303.09175}{{arXiv:2303.09175 [hep-ph]}}.

\bibitem{Belle:2005rte}
S.~Uehara {\em et~al.}, (Belle Collaboration), {\em {Observation of a
  $\chi'_{c2}$ Candidate in $\gamma \gamma \to D \bar{D}$ Production at
  Belle},} \href{http://dx.doi.org/10.1103/PhysRevLett.96.082003}{Phys. Rev.
  Lett. {\bfseries 96} (2006) 082003},
  \href{http://arxiv.org/abs/hep-ex/0512035}{{arXiv:hep-ex/0512035}}.

\bibitem{BaBar:2010jfn}
B.~Aubert {\em et~al.}, (BABAR Collaboration), {\em {Observation of the
  $\chi_{c2}(2P)$ meson in the reaction $\gamma \gamma \to D \bar{D}$ at
  BABAR},} \href{http://dx.doi.org/10.1103/PhysRevD.81.092003}{Phys. Rev. D
  {\bfseries 81} (2010) 092003},
  \href{http://arxiv.org/abs/1002.0281}{{arXiv:1002.0281 [hep-ex]}}.

\bibitem{Guo:2012tv}
F.-K. Guo and U.-G. Meissner, {\em {Where is the $\chi_{c0}(2P)$?},}
  \href{http://dx.doi.org/10.1103/PhysRevD.86.091501}{Phys. Rev. D {\bfseries
  86} (2012) 091501}, \href{http://arxiv.org/abs/1208.1134}{{arXiv:1208.1134
  [hep-ph]}}.

\bibitem{Olsen:2014maa}
S.~L. Olsen, {\em {Is the $X$(3915) the $\chi_{c0}(2P)$?},}
  \href{http://dx.doi.org/10.1103/PhysRevD.91.057501}{Phys. Rev. D {\bfseries
  91} no.~5, (2015) 057501},
  \href{http://arxiv.org/abs/1410.6534}{{arXiv:1410.6534 [hep-ex]}}.

\bibitem{BaBar:2012nxg}
J.~P. Lees {\em et~al.}, (BABAR Collaboration), {\em {Study of $X(3915) \to
  J/\psi \omega$ in two-photon collisions},}
  \href{http://dx.doi.org/10.1103/PhysRevD.86.072002}{Phys. Rev. D {\bfseries
  86} (2012) 072002}, \href{http://arxiv.org/abs/1207.2651}{{arXiv:1207.2651
  [hep-ex]}}.

\bibitem{Belle:2004lle}
K.~Abe {\em et~al.}, (Belle Collaboration), {\em {Observation of a
  near-threshold $\omega J/\psi$ mass enhancement in exclusive $B \to K \omega
  J/\psi$ decays},}
  \href{http://dx.doi.org/10.1103/PhysRevLett.94.182002}{Phys. Rev. Lett.
  {\bfseries 94} (2005) 182002},
  \href{http://arxiv.org/abs/hep-ex/0408126}{{arXiv:hep-ex/0408126}}.

\bibitem{BaBar:2007vxr}
B.~Aubert {\em et~al.}, (BaBar Collaboration), {\em {Observation of $Y(3940)
  \to J/\psi \omega$ in $B \to J/\psi \omega K$ at BABAR},}
  \href{http://dx.doi.org/10.1103/PhysRevLett.101.082001}{Phys. Rev. Lett.
  {\bfseries 101} (2008) 082001},
  \href{http://arxiv.org/abs/0711.2047}{{arXiv:0711.2047 [hep-ex]}}.

\bibitem{LHCb:2020pxc}
R.~Aaij {\em et~al.}, (LHCb Collaboration), {\em {Amplitude analysis of the
  $B^+ \to D^+D^-K^+$ decay},}
  \href{http://dx.doi.org/10.1103/PhysRevD.102.112003}{Phys. Rev. D {\bfseries
  102} (2020) 112003}, \href{http://arxiv.org/abs/2009.00026}{{arXiv:2009.00026
  [hep-ex]}}.

\bibitem{Belle:2017egg}
K.~Chilikin {\em et~al.}, (Belle Collaboration), {\em {Observation of an
  alternative $\chi_{c0}(2P)$ candidate in $e^+ e^- \to J/\psi D \bar{D}$},}
  \href{http://dx.doi.org/10.1103/PhysRevD.95.112003}{Phys. Rev. D {\bfseries
  95} (2017) 112003}, \href{http://arxiv.org/abs/1704.01872}{{arXiv:1704.01872
  [hep-ex]}}.

\bibitem{Zhou:2015uva}
Z.-Y. Zhou, Z.~Xiao, and H.-Q. Zhou, {\em {Could the $X(3915)$ and the
  $X(3930)$ Be the Same Tensor State?},}
  \href{http://dx.doi.org/10.1103/PhysRevLett.115.022001}{Phys. Rev. Lett.
  {\bfseries 115} no.~2, (2015) 022001},
  \href{http://arxiv.org/abs/1501.00879}{{arXiv:1501.00879 [hep-ph]}}.

\bibitem{Ortega:2017qmg}
P.~G. Ortega, J.~Segovia, D.~R. Entem, and F.~Fern{\'a}ndez, {\em {Charmonium
  resonances in the 3.9 GeV/$c^2$ energy region and the $X(3915)/X(3930)$
  puzzle},} \href{http://dx.doi.org/10.1016/j.physletb.2018.01.005}{Phys. Lett.
  B {\bfseries 778} (2018) 1},
  \href{http://arxiv.org/abs/1706.02639}{{arXiv:1706.02639 [hep-ph]}}.

\bibitem{Baru:2017fgv}
V.~Baru, C.~Hanhart, and A.~V. Nefediev, {\em {Can $X(3915)$ be the tensor
  partner of the $X(3872)$?},}
  \href{http://dx.doi.org/10.1007/JHEP06(2017)010}{JHEP {\bfseries 06} (2017)
  010}, \href{http://arxiv.org/abs/1703.01230}{{arXiv:1703.01230 [hep-ph]}}.

\bibitem{LHCb:2019lnr}
R.~Aaij {\em et~al.}, (LHCb Collaboration), {\em {Near-threshold $D\bar{D}$
  spectroscopy and observation of a new charmonium state},}
  \href{http://dx.doi.org/10.1007/JHEP07(2019)035}{JHEP {\bfseries 07} (2019)
  035}, \href{http://arxiv.org/abs/1903.12240}{{arXiv:1903.12240 [hep-ex]}}.

\bibitem{Chen:2012wy}
D.-Y. Chen, J.~He, X.~Liu, and T.~Matsuki, {\em {Does the enhancement observed
  in $\gamma\gamma \to D\bar{D}$ contain two $P$-wave higher charmonia?},}
  \href{http://dx.doi.org/10.1140/epjc/s10052-012-2226-4}{Eur. Phys. J. C
  {\bfseries 72} (2012) 2226},
  \href{http://arxiv.org/abs/1207.3561}{{arXiv:1207.3561 [hep-ph]}}.

\bibitem{Deineka:2021aeu}
O.~Deineka, I.~Danilkin, and M.~Vanderhaeghen, {\em {Dispersive analysis of the
  $\gamma \gamma \to D \bar{D}$ data and the confirmation of the $D \bar{D}$
  bound state},} \href{http://dx.doi.org/10.1016/j.physletb.2022.136982}{Phys.
  Lett. B {\bfseries 827} (2022) 136982},
  \href{http://arxiv.org/abs/2111.15033}{{arXiv:2111.15033 [hep-ph]}}.

\bibitem{Wang:2020elp}
E.~Wang, H.-S. Li, W.-H. Liang, and E.~Oset, {\em {Analysis of the $\gamma
  \gamma \to D \bar{D}$ reaction and the $D \bar{D}$ bound state},}
  \href{http://dx.doi.org/10.1103/PhysRevD.103.054008}{Phys. Rev. D {\bfseries
  103} no.~5, (2021) 054008},
  \href{http://arxiv.org/abs/2010.15431}{{arXiv:2010.15431 [hep-ph]}}.

\bibitem{Ji:2022vdj}
T.~Ji, X.-K. Dong, M.~Albaladejo, M.-L. Du, F.-K. Guo, J.~Nieves, and B.-S.
  Zou, {\em {Understanding the $0^{++}$ and $2^{++}$ charmonium(-like) states
  near 3.9~GeV},} \href{http://dx.doi.org/10.1016/j.scib.2023.02.034}{Sci.
  Bull. {\bfseries 68} (2023) 688},
  \href{http://arxiv.org/abs/2212.00631}{{arXiv:2212.00631 [hep-ph]}}.

\bibitem{Luszczak:2011js}
M.~{\L}uszczak and A.~Szczurek, {\em {Exclusive $D \bar{D}$ meson pair
  production in peripheral ultrarelativistic heavy-ion collisions},}
  \href{http://dx.doi.org/10.1016/j.physletb.2011.04.050}{Phys. Lett. B
  {\bfseries 700} (2011) 116},
  \href{http://arxiv.org/abs/1103.4268}{{arXiv:1103.4268 [nucl-th]}}.

\bibitem{Sobrinho:2024tre}
F.~C. Sobrinho, L.~M. Abreu, C.~A. Bertulani, and F.~S. Navarra, {\em
  {Production of meson molecules in ultraperipheral heavy ion collisons},}
  \href{http://dx.doi.org/10.1103/PhysRevD.110.034037}{Phys. Rev. D {\bfseries
  110} no.~3, (2024) 034037},
  \href{http://arxiv.org/abs/2405.02645}{{arXiv:2405.02645 [hep-ph]}}.

\bibitem{Lebiedowicz:2009pj}
P.~Lebiedowicz and A.~Szczurek, {\em {Exclusive $pp \to pp \pi^{+}\pi^{-}$
  reaction: From the threshold to LHC},}
  \href{http://dx.doi.org/10.1103/PhysRevD.81.036003}{Phys.Rev. {\bfseries D81}
  (2010) 036003},
\href{http://arxiv.org/abs/0912.0190}{{arXiv:0912.0190 [hep-ph]}}.

\bibitem{Lebiedowicz:2018muq}
P.~Lebiedowicz and A.~Szczurek, {\em {Exclusive and semiexclusive production of
  $\mu^+ \mu^-$ pairs with $\Delta$ isobars and other resonances in the final
  state and the size of absorption effects},}
  \href{http://dx.doi.org/10.1103/PhysRevD.98.053007}{Phys. Rev. D {\bfseries
  98} no.~5, (2018) 053007},
  \href{http://arxiv.org/abs/1807.06069}{{arXiv:1807.06069 [hep-ph]}}.

\bibitem{Poppe:1986dq}
M.~Poppe, {\em {Exclusive Hadron Production in Two Photon Reactions},}
\href{http://dx.doi.org/10.1142/S0217751X8600023X}{Int.J.Mod.Phys. {\bfseries
  A1} (1986) 545}.

\bibitem{Pascalutsa:2012pr}
V.~Pascalutsa, V.~Pauk, and M.~Vanderhaeghen, {\em {Light-by-light scattering
  sum rules constraining meson transition form factors},}
  \href{http://dx.doi.org/10.1103/PhysRevD.85.116001}{Phys. Rev. {\bfseries
  D85} (2012) 116001},
\href{http://arxiv.org/abs/1204.0740}{{arXiv:1204.0740 [hep-ph]}}.

\bibitem{Babiarz:2020jkh}
I.~Babiarz, R.~Pasechnik, W.~Sch{\"a}fer, and A.~Szczurek, {\em
  {Hadroproduction of scalar $P$-wave quarkonia in the light-front
  $k_{T}$-factorization approach},}
  \href{http://dx.doi.org/10.1007/JHEP06(2020)101}{JHEP {\bfseries 06} (2020)
  101}, \href{http://arxiv.org/abs/2002.09352}{{arXiv:2002.09352 [hep-ph]}}.

\bibitem{Schuler:1997yw}
G.~A. Schuler, F.~A. Berends, and R.~van Gulik, {\em {Meson photon transition
  form-factors and resonance cross-sections in $e^+ e^-$ collisions},}
  CERN-TH-97-294, \href{http://dx.doi.org/10.1016/S0550-3213(98)00128-X}{Nucl.
  Phys. B {\bfseries 523} (1998) 423--438},
  \href{http://arxiv.org/abs/hep-ph/9710462}{{arXiv:hep-ph/9710462}}.

\bibitem{Babiarz:2024sqw}
I.~Babiarz, R.~Pasechnik, W.~Sch{\"a}fer, and A.~Szczurek, {\em {$\chi_{c2}$
  tensor meson transition form factors in the light front approach},}
  \href{http://dx.doi.org/10.1007/JHEP06(2024)159}{JHEP {\bfseries 06} (2024)
  159}, \href{http://arxiv.org/abs/2402.13910}{{arXiv:2402.13910 [hep-ph]}}.

\bibitem{Li:2021ejv}
Y.~Li, M.~Li, and J.~P. Vary, {\em {Two-photon transitions of charmonia on the
  light front},} \href{http://dx.doi.org/10.1103/PhysRevD.105.L071901}{Phys.
  Rev. D {\bfseries 105} no.~7, (2022) L071901},
  \href{http://arxiv.org/abs/2111.14178}{{arXiv:2111.14178 [hep-ph]}}.

\bibitem{Eichten:2019hbb}
E.~J. Eichten and C.~Quigg, {\em {Quarkonium wave functions at the origin: an
  update},} \href{http://arxiv.org/abs/1904.11542}{{arXiv:1904.11542
  [hep-ph]}}.

\bibitem{Ewerz:2013kda}
C.~Ewerz, M.~Maniatis, and O.~Nachtmann, {\em {A Model for Soft High-Energy
  Scattering: Tensor Pomeron and Vector Odderon},}
  \href{http://dx.doi.org/http://dx.doi.org/10.1016/j.aop.2013.12.001}{Annals
  Phys. {\bfseries 342} (2014) 31},
\href{http://arxiv.org/abs/1309.3478}{{arXiv:1309.3478 [hep-ph]}}.

\bibitem{Wilson:2023anv}
D.~J. Wilson, C.~E. Thomas, J.~J. Dudek, and R.~G. Edwards, (Hadron Spectrum),
  {\em {Charmonium $\chi_{c0}$ and $\chi_{c2}$ resonances in coupled-channel
  scattering from lattice QCD},}
  \href{http://dx.doi.org/10.1103/PhysRevD.109.114503}{Phys. Rev. D {\bfseries
  109} no.~11, (2024) 114503},
  \href{http://arxiv.org/abs/2309.14071}{{arXiv:2309.14071 [hep-lat]}}.

\bibitem{Gui:2018rvv}
L.-C. Gui, L.-S. Lu, Q.-F. L{\"u}, X.-H. Zhong, and Q.~Zhao, {\em {Strong
  decays of higher charmonium states into open-charm meson pairs},}
  \href{http://dx.doi.org/10.1103/PhysRevD.98.016010}{Phys. Rev. D {\bfseries
  98} no.~1, (2018) 016010},
  \href{http://arxiv.org/abs/1801.08791}{{arXiv:1801.08791 [hep-ph]}}.

\bibitem{Wang:2022dfd}
Z.-H. Wang and G.-L. Wang, {\em {Two-body strong decays of the 2P and 3P
  charmonium states},}
  \href{http://dx.doi.org/10.1103/PhysRevD.106.054037}{Phys. Rev. D {\bfseries
  106} no.~5, (2022) 054037},
  \href{http://arxiv.org/abs/2204.08236}{{arXiv:2204.08236 [hep-ph]}}.

\bibitem{Eichten:2004uh}
E.~J. Eichten, K.~Lane, and C.~Quigg, {\em {Charmonium levels near threshold
  and the narrow state $X(3872) \to \pi^{+}\pi^{-}J/\psi$},}
  \href{http://dx.doi.org/10.1103/PhysRevD.69.094019}{Phys. Rev. D {\bfseries
  69} (2004) 094019},
  \href{http://arxiv.org/abs/hep-ph/0401210}{{arXiv:hep-ph/0401210}}.

\bibitem{Man:2024mvl}
Z.-L. Man, C.-R. Shu, Y.-R. Liu, and H. Chen,
{\em {Charmonium states in a coupled-channel model},}
  \href{http://dx.doi.org/10.1140/epjc/s10052-024-13132-7}{Eur. Phys. J. C {\bfseries 84} no.~8, (2024) 810},
  \href{https://doi.org/10.48550/arXiv.2402.02765}{{arXiv:2402.02765 [hep-ph]}}.
  
\bibitem{Zhou:2017dwj}
Z.-Y. Zhou and Z. Xiao,
{\em {Understanding $X(3862)$, $X(3872)$, and $X(3930)$ in a Friedrichs-model-like scheme},}
  \href{http://dx.doi.org/10.1103/PhysRevD.96.054031}{Phys. Rev. D {\bfseries
  96} (2017) 5},
  \href{https://doi.org/10.48550/arXiv.1704.04438}{{arXiv:hep-ph/1704.04438}}. 
  [Erratum: Phys.Rev.D 96, 099905(E) (2017)].

\bibitem{Yu:2017bsj}
G.-L. Yu, Z.-G. Wang, and Z.-Y. Li,
{\em {Analysis of the charmonium-like states $X^{*}(3860)$, $X(3872)$, $X(3915)$, $X(3930)$ and $X(3940)$ according to their strong decay behaviors},}
  \href{http://dx.doi.org/10.1088/1674-1137/42/4/043107}{Chin. Phys. C {\bfseries
  42} (2018) 4},
  \href{https://doi.org/10.48550/arXiv.1704.06763}{{arXiv:hep-ph/1704.06763}}.

\bibitem{Gao:2025tob}
X.-L. Gao, J.-X. Cui, Y.-H. Zhou, and Z.-Y. Zhou,
{\em {Updated analysis of charmonium states in a relativized quark potential model},}
\href{https://doi.org/10.48550/arXiv.2504.14575}{{arXiv:hep-ph/2504.14575}}.

\bibitem{Jaus:1996np}
W.~Jaus, {\em {Semileptonic, radiative, and pionic decays of $B$, $B^*$ and
  $D$, $D^*$ mesons},} \href{http://dx.doi.org/10.1103/PhysRevD.53.1349}{Phys.
  Rev. D {\bfseries 53} (1996) 1349}. [Erratum: Phys.Rev.D 54, 5904 (1996)].

\bibitem{Rosner:2013sha}
J.~L. Rosner, {\em {Hadronic and radiative $D^*$ widths},}
  \href{http://dx.doi.org/10.1103/PhysRevD.88.034034}{Phys. Rev. D {\bfseries
  88} no.~3, (2013) 034034},
  \href{http://arxiv.org/abs/1307.2550}{{arXiv:1307.2550 [hep-ph]}}.

\bibitem{Guo:2019qcn}
F.-K. Guo, {\em {Novel Method for Precisely Measuring the $X(3872)$ Mass},}
  \href{http://dx.doi.org/10.1103/PhysRevLett.122.202002}{Phys. Rev. Lett.
  {\bfseries 122} no.~20, (2019) 202002},
  \href{http://arxiv.org/abs/1902.11221}{{arXiv:1902.11221 [hep-ph]}}.

\bibitem{Jia:2024imm}
M.~Jia, W.~Li, S.-Y. Pei, X.-Y. Du, G.-Z. Ning, and G.-L. Wang, {\em
  {Relativistic effects in the strong and electromagnetic decays of the $D^*$
  meson},} \href{http://dx.doi.org/10.1140/epjc/s10052-025-13974-9}{Eur. Phys.
  J. C {\bfseries 85} no.~3, (2025) 282},
  \href{http://arxiv.org/abs/2412.10775}{{arXiv:2412.10775 [hep-ph]}}.

\bibitem{Cao:2024nxm}
X.-H. Cao, M.-L. Du, and F.-K. Guo, {\em {Photoproduction of the $X(3872)$
  beyond vector meson dominance: the open-charm coupled-channel mechanism},}
  \href{http://dx.doi.org/10.1088/1361-6471/ad6fbc}{J. Phys. G {\bfseries 51}
  no.~10, (2024) 105002},
  \href{http://arxiv.org/abs/2401.16112}{{arXiv:2401.16112 [hep-ph]}}.

\bibitem{Lebiedowicz:2019jru}
P.~Lebiedowicz, O.~Nachtmann, and A.~Szczurek, {\em {Central exclusive
  diffractive production of $K^{+} K^{-} K^{+} K^{-}$ via the intermediate
  $\phi \phi$ state in proton-proton collisions},}
  \href{http://dx.doi.org/10.1103/PhysRevD.99.094034}{Phys. Rev. {\bfseries
  D99} no.~9, (2019) 094034},
\href{http://arxiv.org/abs/1901.11490}{{arXiv:1901.11490 [hep-ph]}}.

\bibitem{Lebiedowicz:2021pzd}
P.~Lebiedowicz, {\em {Study of the exclusive reaction $pp \to pp K^{*0}
  \bar{K}^{*0}$: $f_{2}(1950)$ resonance versus diffractive continuum},}
  \href{http://dx.doi.org/10.1103/PhysRevD.103.054039}{Phys. Rev. D {\bfseries
  103} no.~5, (2021) 054039},
  \href{http://arxiv.org/abs/2102.13029}{{arXiv:2102.13029 [hep-ph]}}.

\bibitem{Brisudova:1999ut}
M.~M. Brisudov{\'a}, L.~Burakovsky, and J.~T. Goldman, {\em {Effective
  functional form of Regge trajectories},}
  \href{http://dx.doi.org/10.1103/PhysRevD.61.054013}{Phys. Rev. D {\bfseries
  61} (2000) 054013},
  \href{http://arxiv.org/abs/hep-ph/9906293}{{arXiv:hep-ph/9906293}}.

\bibitem{Lebiedowicz:2015cea}
P.~Lebiedowicz and A.~Szczurek, {\em {Exclusive production of heavy charged
  Higgs boson pairs in the $p p \to p p H^+ H^-$ reaction at the LHC and a
  future circular collider},}
  \href{http://dx.doi.org/10.1103/PhysRevD.91.095008}{Phys. Rev. D {\bfseries
  91} (2015) 095008}, \href{http://arxiv.org/abs/1502.03323}{{arXiv:1502.03323
  [hep-ph]}}.

\bibitem{Klusek-Gawenda:2017lgt}
M.~K{\l}usek-Gawenda, P.~Lebiedowicz, O.~Nachtmann, and A.~Szczurek, {\em {From
  the $\gamma \gamma \to p \bar{p}$ reaction to the production of $p \bar{p}$
  pairs in ultraperipheral ultrarelativistic heavy-ion collisions at the LHC},}
  \href{http://dx.doi.org/10.1103/PhysRevD.96.094029}{Phys. Rev. {\bfseries
  D96} no.~9, (2017) 094029},
\href{http://arxiv.org/abs/1708.09836}{{arXiv:1708.09836 [hep-ph]}}.

\bibitem{Swanson:2006st}
E. S. Swanson, {\em {The new heavy mesons: A status report},}
  \href{http://dx.doi.org/10.1016/j.physrep.2006.04.003}{Phys. Rept. {\bfseries
  429} (2006) 243},
\href{https://doi.org/10.48550/arXiv.hep-ph/0601110}{{arXiv:0601110 [hep-ph]}}.
  
\end{thebibliography}

\end{document}